\documentclass[11pt]{article}
\usepackage{amsfonts}
\usepackage{amsmath}
\usepackage{amssymb}
\usepackage{bbm,bm}
\usepackage{braket} %luis
\usepackage{empheq} %luis
\usepackage{slashed} %luis
\usepackage{tabularx, booktabs}
\usepackage{cite}
\allowdisplaybreaks[1]

\usepackage{comment}
\usepackage{float} 
\usepackage[normalem]{ulem} % \sout

\usepackage[
  colorlinks=true,
  linkcolor={red!50!black},
  citecolor={blue!50!black},
  filecolor=black,
  urlcolor=black,
  breaklinks=true
  ]{hyperref}
\usepackage{orcidlink}

\newcommand{\Dpath}{\mathcal{D}}
\newcommand{\tr}{\mathrm{Tr}}
\newcommand{\imag}[1]{\mathrm{Im}\left\{ #1 \right\}}

\newcolumntype{Y}{>{\centering\arraybackslash}X}

\makeatletter \@addtoreset{equation}{section} \makeatother
\renewcommand{\theequation}{\arabic{section}.\arabic{equation}}
\makeatletter
\renewcommand\section{\@startsection{section}{1}{\z@}%
  {-5.5ex \@plus -1ex \@minus -.2ex}% bfr-skip
  {2.3ex \@plus.2ex}%
  {\normalfont\large\bfseries}}
\renewcommand\subsection{\@startsection{subsection}{2}{\z@}%
  {-3.25ex\@plus -1ex \@minus -.2ex}%
  {1.5ex \@plus .2ex}%
  {\normalfont\normalsize\bfseries}}
\renewcommand\thesection{\@arabic\c@section}
\renewcommand\thesubsection{\thesection.\@arabic\c@subsection}
\renewcommand{\@seccntformat}[1]{%
  \csname the#1\endcsname.\hspace{1.0em}}
\makeatother

\begin{document}

\flushbottom

\begin{titlepage}

\begin{centering}

\vfill

{\large{\bf
    A consistent (3D) EFT expansion of the bubble nucleation rate. \\
    Part I: scalar theories 
  }
}

\vspace{0.8cm}

\renewcommand{\thefootnote}{\fnsymbol{footnote}}
Luis Gil%
\orcidlink{0009-0007-4502-7521}%
\footnote{lgil@ugr.es}%
and
Javier López Miras%
\orcidlink{0000-0002-7481-3827}%
\footnote{jlmiras@ugr.es}%

\vspace{0.8cm}

{\em
Departamento de F\'isica Te\'orica y del Cosmos, Universidad de Granada,\\
Campus de Fuentenueva, E–18071 Granada, Spain\\}

\vspace*{0.8cm}

\mbox{\bf Abstract}

\end{centering}

\vspace*{0.3cm}

\noindent
Precise predictions of bubble nucleation are essential for connecting cosmological phase transitions to their potential observational signatures. In this first paper of a two-part series, we develop a general, fully perturbative framework for incorporating higher-dimensional operators (HDOs) into nucleation calculations in scalar high-temperature effective field theories. Our approach consistently includes their effects on the bounce and the fluctuations around it, including interactions with higher derivatives. We prove that the resulting statistical rate is invariant under field redefinitions, and establish sufficient conditions for the existence of the perturbative bounce solution in the presence of HDOs. We illustrate the framework by applying it to a model containing a scalar and a fermion, where we find modest HDO corrections in its regime of moderately strong transitions. We thus extend the systematic study of HDO effects beyond critical thermodynamics and provide a controlled method for determining their impact on bubble nucleation.

\vfill
\end{titlepage}

{\hypersetup{hidelinks}
\tableofcontents
}
\clearpage

\renewcommand{\thefootnote}{\arabic{footnote}}
\setcounter{footnote}{0}

\section{Introduction}

With the existing and upcoming generation of gravitational wave (GW) interferometers \cite{Harry:2006fi,Kawamura:2006up,Ruan:2018tsw,LIGOScientific:2014pky,NANOGrav:2020bcs}, there has arisen a great interest to precisely characterize the many potential sources of GWs in the universe. Cosmological phase transitions (PTs) \cite{Caprini:2019egz} have particularly drawn the attention of both cosmologists and particle physicists in recent years. The reason is clear: the Standard Model (SM) does not predict a first-order PT \cite{Kajantie:1996mn, Gurtler:1997hr, Csikor:1998eu}, so any observation of a GW background compatible with a first-order PT would constitute compelling evidence for physics beyond the SM.

In order to bridge the gap between future experimental data and the vast new-physics landscape, it is crucial to be able to produce precise predictions of the PT (thermo)dynamics. As of today, the most reliable approach to the study of thermal PTs is through effective field theory (EFT) methods, specifically, high-temperature ($T$) dimensional reduction (DR) \cite{Ginsparg:1980ef,Appelquist:1981vg,Kajantie:1995dw}. In DR one integrates out the heavy thermal modes \cite{Matsubara:1955ws}, resulting in a three-dimensional EFT (3D EFT) for the bosonic modes which are dynamical at the scale at which the PT occurs. The advantages of the widely extended (see \textit{e.g.} \cite{Brauner:2016fla,Andersen:2017ika,Niemi:2018asa,Gorda:2018hvi,Kainulainen:2019kyp,Croon:2020cgk,Gould:2019qek,Niemi:2020hto,Gould:2021ccf,Gould:2021dzl,Schicho:2021gca,Niemi:2021qvp,Camargo-Molina:2021zgz,Niemi:2022bjg,Ekstedt:2022ceo,Gould:2022ran,Ekstedt:2022zro,Biondini:2022ggt,Schicho:2022wty,Lofgren:2021ogg,Gould:2023jbz,Kierkla:2023von,Aarts:2023vsf,Niemi:2024axp,Chala:2024xll,Qin:2024idc,Gould:2024jjt,Chakrabortty:2024wto,Niemi:2024vzw,Kierkla:2025qyz,Bhatnagar:2025jhh,Bernardo:2025vkz,Chala:2025aiz,Zhu:2025pht,Chala:2025oul,Li:2025kyo,Annala:2025aci,Navarrete:2025yxy,Chala:2025xlk,Chala:2025cya,Biekotter:2025npc}) 3D EFT approach are: reduced uncertainties from consistent power counting for perturbative expansions \cite{Croon:2020cgk,Kierkla:2026bnm}, manifest gauge and renormalization scale invariance \cite{Bernardo:2025vkz,Chala:2025aiz}, and feasibility of nonperturbative lattice studies \cite{Nielsen:1981hk}. The recent publication of automated matching tools \cite{Ekstedt:2022bff,Fuentes-Martin:2026bhr,Bernardo:2026nyq} also makes this approach more accessible to the community.

Nevertheless, it was recently shown that in the strong PT regime the hierarchical ratio of scales underlying the effective expansion in DR becomes large, eventually leading to the breakdown of EFTs \cite{Chala:2024xll,Chala:2025oul,Chala:2025aiz,Bernardo:2025vkz,Chala:2025cya,Biekotter:2025npc,Bernardo:2026whs}. Within perturbation theory, there have appeared two responses in the literature: (a) the development of alternatives which do not rely on dimensional reduction~\cite{Navarrete:2025yxy,Berthiaume:2026dkt}, or (b) the extension of the usual 3D EFT expansion to increasingly high orders, pushing its limit of validity~\cite{Chala:2024xll,Chala:2025oul,Chala:2025aiz,Bernardo:2025vkz,Chala:2025cya,Biekotter:2025npc,Bernardo:2026whs,Gould:2026rqz}. While the first approach promises to address the underlying problem, it has so far only been successfully applied to the critical thermodynamics of gauge-less theories. The second approach, which we follow in this series of works, is valid for any theory, and it has led to the precise determination of the thermodynamics of scalar-fermion \cite{Chala:2024xll,Chala:2025oul}, gauge theories \cite{Bernardo:2025vkz,Bernardo:2026whs} and even the full dimension-6 SMEFT \cite{Chala:2025aiz,Chala:2025xlk}. This has set rigorous limits on the validity of 3D EFTs, and it has also shown that higher-dimensional operators (HDOs) significantly alter the predictions for moderately strong PTs within the regime of validity of the EFT. 

Still, by the time we started working on this project, neither (a) nor (b) had so far tackled the computation of the nucleation rate. This observable sets the $T$ scale for the effective start of PTs, and is thus an essential input for the simulations of bubble dynamics \cite{Caprini:2019egz}. Its computation amounts to, among others, the evaluation of fluctuation determinants on a non-homogeneous field configuration, the \textit{bounce} \cite{Coleman:1977py}. While there exist useful public codes for the computation of the bounce (\textit{e.g.} \cite{Wainwright:2011kj,Guada:2020xnz}), and the subsequent perturbative evaluation of the determinants \cite{Ekstedt:2023sqc}, they both require the input 3D EFT to be superrenormalizable (or all HDOs to be resummed in a potential). Very recently, however, the authors of \cite{Bernardo:2026tki} estimated the corrections from dimension-6 3D operators on the nucleation rate in the classically-conformal Abelian Higgs model. They found that corrections due to these HDOs are small in this scenario, confirming the expectation \cite{Levi:2022bzt,Kierkla:2023von} that the high-$T$ expansion works well for nucleation in the supercooled scenario. 

In this first paper of a two-part series, we focus on scalar fields to study the impact of HDOs on the nucleation rate from a complementary perspective. Rather than incorporating these operators directly into the bounce equations and fluctuation operators, we develop a systematic perturbative expansion in their Wilson coefficients in the spirit of~\cite{Chala:2024xll}, based on analogous techniques for the computation of higher-loop corrections to the nucleation rate \cite{Ekstedt:2022tqk}. This distinction is particularly relevant within the EFT framework, where physical predictions must be organized consistently order by order in the expansion. Retaining HDOs without expanding in their coefficients amounts to a partial resummation of higher-order contributions, which are not necessarily accompanied by all other terms entering at the same order in the EFT power counting. Our approach instead isolates the complete correction at the first order, providing a well-defined prescription for assessing the impact of higher EFT orders, one by one, on the nucleation rate.

The advantages of this perturbative EFT treatment are manifold. We establish field-redefinition invariance explicitly, first analytically and then numerically in an example, showing that the equilibrium piece of the nucleation rate is operator-basis independent. We additionally provide a proof of the existence of the bounce solution computed this way under general perturbations due to HDOs. From a computational perspective, our approach also differs substantially from that of \cite{Bernardo:2026tki}. Whereas the latter relies on the Gel'fand-Yaglom method, requiring the fluctuation operators to admit a suitable differential operator representation, we formulate the perturbative corrections in terms of functional traces. This allows us to express the contributions of HDOs as vertex insertions into the leading-order propagators, without the need to solve the modified fluctuation equations. Besides constituting an independent method for evaluating the nucleation rate, this formulation makes the EFT expansion manifest and allows for the contributions of the different operators to be identified separately. Furthermore, it is not tied to the particular form of the fluctuation operators required by the implementation of ref.~\cite{Bernardo:2026tki}, opening the way to applications involving arbitrary HDOs.

We apply our framework to compute the higher-dimensional corrections to the nucleation rate in the high-$T$ limit of the scalar-Yukawa model \cite{Gould:2023jbz,Chala:2024xll,Chala:2025oul}. Our results show that in its moderately-strong transition regime, HDOs only have a mild effect on the nucleation temperature, similar to what was found in the classically conformal Abelian Higgs model~\cite{Bernardo:2026tki}. Since our approach is linear in HDOs, we also quantify the contribution from each operator to the nucleation rate, and show that potential terms are more relevant. This supports the idea that a more appropriate high-$T$ power counting should resum field insertions, while keeping an expansion in derivatives, as recently proposed in \cite{Gould:2026rqz}.

\section{Overview}
\label{sec:overview}

In Langer's framework of classical nucleation theory \cite{Langer:1969bc}, the thermal nucleation (or escape) rate is given by
\begin{equation}
    \Gamma = \Gamma_\mathrm{dyn} \times \Gamma_\mathrm{stat}\,.
    \label{eq:gamma}
\end{equation}
Here, the first factor $\Gamma_\mathrm{dyn}$ captures the non-equilibrium and dissipative effects due to interactions with the environment. The second, $\Gamma_\mathrm{stat}$, is the equilibrium piece, and it is given by a Boltzmann integral over all fluctuations on the bounce background $\varphi_\mathrm{b}$.

The thermal nucleation rate has been proven to factorize into these two factors to all orders assuming Langevin dynamics \cite{Ekstedt:2022tqk}. A first-principles derivation of the thermal nucleation dynamics, along with a careful examination of the assumptions on which formula \eqref{eq:gamma} rests was recently presented in \cite{Gould:2026mtl}.

The statistical piece of the thermal nucleation process is manifestly real to all orders \cite{Ekstedt:2022tqk}, as opposed to the quantum tunneling rate. Still, this statistical rate can be related to the analytic continuation of the exponential of the quantum effective action $S_\mathrm{eff}$ evaluated at the bounce configuration $\boldsymbol\varphi_\mathbf{b}$ \cite{Coleman:1977py}, normalized by the false vacuum (FV), $\boldsymbol\varphi_\mathbf{FV}$ \cite{Gould:2021ccf}. As a path integral:
\begin{equation}
    \Gamma_\mathrm{stat} = \left|\imag{e^{-(S_\mathrm{eff}[\boldsymbol\varphi_\mathbf{b}] - S_\mathrm{eff}[\boldsymbol\varphi_\mathbf{FV}])}}\right| \equiv \left|\imag{\frac{\int [\Dpath \eta]\, e^{-S[\boldsymbol\varphi_\mathbf{b} + \eta]}}{\int [\Dpath \xi]\, e^{-S[\boldsymbol\varphi_\mathbf{FV} + \xi]}}}\right|\,,
\end{equation}
where $[\Dpath \eta]$ ($[\Dpath \xi]$) denotes the integration over all fluctuation modes on the bounce (FV) background. Furthermore, the bounce (FV) state is given by $\boldsymbol\varphi_\mathbf{b(FV)} = (\varphi_\mathrm{b(FV)}, \mathbf{0})$, representing the configuration in which the scalar field sits at the bounce (FV) solution and all the other background fields vanish. In a high-$T$ approximation, $S[\boldsymbol\phi]$ corresponds to the 3D effective action for all bosonic, zero Matsubara modes, that are not heavy compared to the nucleation scale. In this equilibrium EFT, all $T$ dependence and quantum corrections from heavier Matsubara modes are encoded in the Wilson coefficients of effective operators.

In the expression above, the FV state is the higher local minimum of the potential $V$ of the scalar that drives the PT,
\begin{equation}
    \frac{d V}{d \phi}\Biggr|_{\boldsymbol{\varphi}_\mathbf{FV} } = 0\,.
\end{equation}
On the other hand, the bounce is an $O(3)$-symmetric configuration of the transitioning scalar that is a saddle point of $S$. It is given by the equation of motion (EOM) \cite{Coleman:1977py}
\begin{equation}
    \frac{\delta S}{\delta \phi} \Biggr|_{\boldsymbol{\varphi}_\mathbf{b}} = 0 \,; \qquad \partial_r \varphi_\mathrm{b}(r) \bigr|_{r=0} = 0\,,\quad \lim_{r\to\infty} \varphi_\mathrm{b}(r) = \varphi_\mathrm{FV}\,.
    \label{eq:Seff bounce}
\end{equation}

At quadratic order in fluctuations on the background configurations,
\begin{equation}
    S[\boldsymbol\varphi + \hbar^{1/2} \boldsymbol\eta] = S[\boldsymbol\varphi] + \hbar^{1/2} \eta_I S'[\boldsymbol\varphi]_I + \frac{\hbar}{2} \eta_I S''[\boldsymbol\varphi]_{IJ} \eta_J + \mathcal{O}(\hbar^{3/2})\,,
\end{equation}
where we have introduced the shorthand notation
\begin{equation}
    S'[\boldsymbol\varphi]_{I} \equiv \frac{\delta S}{\delta \eta_I}\Biggr|_{\boldsymbol\eta=\boldsymbol\varphi} \,,\qquad S''[\boldsymbol\varphi]_{IJ} \equiv \frac{\delta^2 S}{\delta \eta_I \delta \eta_J} \Biggr|_{\boldsymbol\eta=\boldsymbol\varphi}\,,
\end{equation}
having employed DeWitt notation where the capital indices $I, J$ indicate both space and internal degrees of freedom, so $A_I B_I \equiv \int_x A_a(x) B_a(x)$, with $\int_x \equiv \int d^D x$. Note that the single functional derivative term vanishes for both $\varphi=\boldsymbol\varphi_\mathbf{b}, \boldsymbol\varphi_\mathbf{FV}$.

Since $S''[\boldsymbol\varphi]$ is a self-adjoint operator, its (Gaussian) path integral can be expressed as a functional determinant. However, in order to do this, one has to treat separately all modes of $S''[\boldsymbol\varphi_\mathbf{b}]$ with a vanishing eigenvalue (zero modes), that make $S''[\boldsymbol\varphi_\mathbf{b}]$ noninvertible. This can be achieved via the method of collective coordinates \cite{Andreassen:2016cvx} (see section \ref{sec:zero modes}), leaving us with the following generic expression for the NLO statistical rate (after setting $\hbar=1$):
\begin{equation}
    \Gamma_\mathrm{stat} \thickapprox \mathcal{V} |\det J|\, e^{-(S[\boldsymbol\varphi_\mathbf{b}] - S[\boldsymbol\varphi_\mathbf{FV}])} \left|\frac{\det (S''[\boldsymbol\varphi_\mathbf{FV}]_{IJ})}{{\det}' (S''[\boldsymbol\varphi_\mathbf{b}]_{IJ})} \right|^{1/2} \,.
    \label{eq:Astat}
\end{equation}
Here ${\det}'$ denotes that zero modes are not included in the determinant; $\mathcal{V}$ is a volume factor and $J$ is the Jacobian matrix of the change to collective coordinates.

% The determinants in this expression will factorize for all degrees of freedom that do not mix at quadratic level in the action. This will typically yield separate determinant ratios for scalars, mixed gauge-Goldstone fields and ghosts. Importantly, the exact form of this mixing will depend on the choice of gauge fixing (see e.g. \cite{Isidori:2001bm,Bernardo:2026tki}).

On the other hand, the dynamical rate is not exponential, and for an effective damping coefficient $\eta$ and at LO it is given by \cite{Langer:1969bc,Ekstedt:2022tqk}
\begin{equation}
    \Gamma_\mathrm{dyn} \thickapprox \frac{1}{2\pi} \left( \sqrt{|\lambda_-| + \frac{\eta^2}{4}} - \frac{\eta}{2} \right)\,,
    \label{eq:dynamical rate}
\end{equation}
where $\lambda_-$ is the negative eigenvalue of $S''[\boldsymbol\varphi_\mathbf{b}]$. While $\lambda_-$ can be determined from the equilibrium theory, $\Gamma_\mathrm{dyn}$ is a real-time observable and must be computed independently (see \textit{e.g.} \cite{Ekstedt:2022tqk, Gould:2026mtl}). In this work we are only interested in higher-order corrections within the equilibrium theory, so we will ignore the dynamical rate throughout. 

\section{Expansion in higher EFT orders}

Our goal is to determine the NLO nucleation rate in a scalar 3D EFT which contains HDOs beyond the superrenormalizable level. The 3D EFT action is formally expressed as a power series
\begin{equation}
    S = \sum_{k=0}^N \epsilon^k S_k
\end{equation}
where $S_0$ contains all superrenormalizable operators and $S_k, k>0$, all higher-dimensional terms\footnote{Note that one may also choose an expansion parameter so that parts of the superrenormalizable sector are also included into $S_k$, with $k>0$. Also, importantly, our perturbative approach can in principle be applied to other EFTs where the power counting is not strictly in operator dimensions. For example, in theories with a power counting for derivatives of the nucleating field, but not in powers of the field itself (see \cite{Gould:2026rqz}).} suppressed by a formal perturbative parameter $\epsilon$.

In practice, we will only consider the zeroth and first order in this expansion, \textit{i.e.} \linebreak $S = S_0 + \epsilon S_1$. We will assume that the latter corresponds to dimension-6 operators (in 4D units), which are naturally generated in DR at high-enough orders in the coupling expansion. With this, we will determine an expression for $\Gamma_\mathrm{stat}$ that is exact to $\mathcal{O}(\epsilon)$, consistent with the 3D EFT truncation. We will also show how this perturbative treatment allows for computing corrections to $\Gamma_\mathrm{stat}$ involving arbitrary HDOs, in particular those with more than two derivatives, thus bypassing the restrictions of the existing methods and codes for the evaluation of nucleation rates \cite{Wainwright:2011kj,Ekstedt:2023sqc,Bernardo:2026tki}.

In this first work, we present a pedagogical derivation for the scalar case in $D$ spatial dimensions, and leave the extension to gauge theories to a forthcoming publication \cite{Gil:2026tbr}.

\subsection{Higher-order corrections to the bounce}

Ref.~\cite{Chala:2024xll} first discussed perturbative modification of the bounce solution due to HDOs at any order (see appendix B therein). Here we summarize those results at $\mathcal{O}(\epsilon)$.

Let $S[\phi] = S_0[\phi] + \epsilon S_1[\phi]$ be the action of a single real scalar field in $D$-dimensional Euclidean space, with
\begin{equation}
    S_0[\phi]= \int_x \left[ \frac{1}{2} (\partial_i \phi)^2 + V_0(\phi)  \right] \,.   
\end{equation}
Assume that $V_0(\phi)$ has a minimum at $\phi=\varphi_\mathrm{FV}$ that we center at the origin, \textit{i.e.} $\varphi_\mathrm{FV}=0$. Also, let $\varphi_\mathrm{b}(x) = \varphi_0(x) + \epsilon \varphi_1(x)$ be a classical solution to the EOM of $S$, with $O(D)$ symmetry and boundary conditions $\dot\varphi_\mathrm{b}(0)=\varphi_\mathrm{b}(\infty)=0$, where $\dot\varphi_\mathrm{b}(r) \equiv \partial_r \varphi_\mathrm{b}(r)$ and $r\equiv |x|$. Namely, $\varphi_\mathrm{b}(r)$ is a bounce solution to $S$ (see eq.~\eqref{eq:Seff bounce}). We assume that $V_0(\phi)$ is such that the bounce configuration exists (we refer to appendix~\ref{app:existence} for more details).

The EOM for $\varphi_{\rm b}$ reads 
\begin{equation}
    0 = S'_0[\varphi_0] + \epsilon \left( S'_1[\varphi_0] + S''_0[\varphi_0] \varphi_1 \right) + \mathcal{O}(\epsilon^2)\,.
\end{equation}
For this equation to hold true, each order must vanish separately. Hence, $\varphi_0$ satisfies
\begin{equation}
    S'_0[\varphi_0] = 0  \quad\implies\quad
    \partial^2 \varphi_0 - V_0'(\varphi_0) = 0
    \,,
    \label{eq:diffeq phi0}
\end{equation}
\noindent
where $\partial^2 = \partial_i \partial_i$ and $V'_0(\varphi_0) \equiv d V_0/d\phi \bigr|_{\varphi_0}$. On the other hand, $\varphi_1$ is given by the following second order partial differential equation:
\begin{equation}
   S''_0[\varphi_0] \varphi_1 = -S'_1[\varphi_0] \quad\implies
  \quad \left(\partial^2 - V_0''(\varphi_0(r)) \right) \varphi_1(r) = S'_1[\varphi_0(r)] \,.
    \label{eq:diffeq phi1}
\end{equation}

The boundary conditions must also be satisfied order by order in $\epsilon$. This means that $\varphi_0(r)$ is the bounce solution to $S_0$, and $\varphi_1(r)$ solves eq.~\eqref{eq:diffeq phi1} with boundary conditions
\begin{equation}
    \dot\varphi_1(0)=\varphi_1(\infty)=0 \,.
    \label{eq:bcs phi1}
\end{equation}

The false vacuum solution must be analogously solved perturbatively. However, after setting $\varphi_{\rm FV, 0}=0$, it follows $\varphi_{\rm FV,1} = 0$, as long as there is no tadpole operator in $S_1[\phi]$.

While the existence of $\varphi_0$ has been proven for actions with a generic kinetic plus potential structure \cite{Coleman:1977py}, we are not aware of a proof of existence for bounce solutions $\varphi$ to the full EOM in the presence of arbitrary HDOs. In our perturbative approach, we are interested in determining whether $\varphi_1$ exists and whether it is unique. In appendix \ref{app:existence} we collect results proving that the existence and uniqueness of $\varphi_0$ implies the same for $\varphi_1$, and for any higher-order correction.

\subsection{Higher-order corrections to the statistical rate}

As we saw in section \ref{sec:overview}, the exponential of the quantum effective action at a background $\varphi$ can be approximated as
\begin{equation}
    e^{-S_\mathrm{eff}[\varphi]} \thickapprox e^{-S[\varphi]} \int \Dpath \eta\, e^{-\frac{1}{2} \eta S''[\varphi]\eta}\,.
    \label{eq:Seff}
\end{equation}
Taking $\varphi$ to be a stationary point of the action, it can be expanded to $\mathcal{O}(\epsilon)$ as
\begin{align}
    e^{-S_\mathrm{eff}[\varphi]} &\thickapprox e^{-S_0[\varphi_0]} \left(1 - \epsilon S_1[\varphi_0]\right) \int \Dpath \eta\, e^{-\frac{1}{2} \eta S_0''[\varphi_0]\eta} \left[ 1 - \frac{1}{2} \epsilon \eta \left( S_1''[\varphi_0] + S_0'''[\varphi_0] \varphi_1 \right)\eta \right] \nonumber\\
    &\thickapprox e^{-S_0[\varphi_0]} \int \Dpath \eta\, e^{-\frac{1}{2} \eta S_0''[\varphi_0]\eta} \times \nonumber\\
    & \hphantom{=} \left\{ 1 - \epsilon \left[ S_1[\varphi_0] + \frac{1}{2} \frac{\int \Dpath \eta\, e^{-\frac{1}{2} \eta S_0''[\varphi_0]\eta} \eta \left( S_1''[\varphi_0] + S_0'''[\varphi_0] \varphi_1 \right) \eta}{\int \Dpath \eta\, e^{-\frac{1}{2} \eta S_0''[\varphi_0]\eta}} \right] \right\}\,.
    \label{eq:Seff expanded}
\end{align}

Before rewriting this expression as a combination of functional determinants and traces, we must watch out for possible zero modes of $S''_0[\varphi_0]$. We address this in what follows.

\subsubsection{Translational zero modes}
\label{sec:zero modes}

By taking the gradient of the EOM of the zeroth-order bounce $\varphi_0$,
\begin{equation}
    S'_0[\varphi_0] = 0 \quad \implies \quad S''_0[\varphi_0] \partial_i \varphi_0 = 0\,,
\end{equation}
we can see that for modes $\eta = \partial_i \varphi_0 \neq 0\,,  i=1,\dots,D$, the integrals in eq.~\eqref{eq:Seff expanded} diverge. For the homogeneous FV, $\partial_i \varphi_\mathrm{FV}=0$, but for $\varphi_\mathrm{b}$ this gradient is non-zero, so in this second case the zero modes of $S''_0[\varphi_0]$ must be treated separately in the path integral.

Since $S''_0[\varphi_0]$ is a self-adjoint operator in the space of square-integrable functions $L^2(\mathbb{R}^D)$, the spectral theorem states that its eigenfunctions (modes) form an orthonormal basis of $L^2(\mathbb{R}^D)$. Also, since the zero modes lie in the discrete part of the spectrum, let us consider integration in a finite space volume $\mathcal{V}(\mathbb{R}^D)$ to avoid keeping track of the continuum modes.

The perturbative eigenvalue equations through $\mathcal{O}(\epsilon)$ are
\begin{equation}
    \begin{cases}
        &S''_0[\varphi_0] \psi_n^{(0)} = \lambda_n^{(0)} \psi_n^{(0)} \\
        &S''_0[\varphi_0] \psi_n^{(1)} + (S_1''[\varphi_0] + S_0'''[\varphi_0] \varphi_1) \psi_n^{(0)} = \lambda_n^{(0)} \psi_n^{(1)} + \lambda_n^{(1)} \psi_n^{(0)}
    \end{cases}\,.
    \label{eq:eigenvalue equations}
\end{equation}
We order the modes so that  $\psi_1^{(0)}, \ldots, \psi_D^{(0)}$ correspond to the zero modes of $S''_0[\varphi_0]$, with $\lambda_1^{(0)} = \ldots = \lambda_D^{(0)} = 0$ and the rest correspond to the single \cite{Coleman:1977th} negative eigenvalue, and all positive eigenvalues, which are ordered starting from $n=D+1$. We assume that $S_0''[\varphi_0]$ has only the translational zero modes, which is true upon uniqueness of the bounce solution (appendix~\ref{app:existence}). Otherwise, the remaining zero modes should be removed as well (see \textit{e.g.}~\cite{Ekstedt:2023sqc}).

We may write any fluctuation over the bounce as a linear combination of modes of $S''_0[\varphi_0]$ with coordinates $\{a_n\}$ in field space. In order to handle the divergent zero mode fluctuations, we change to a basis of collective coordinates $\{b_i,c_n\}$ for some $b \in \mathbb{R}^D$ \cite{Bhattacharya:2024chz}. Namely,
\begin{equation}
    \eta(x) = \sum_{n \geq 1} a_n \psi_n^{(0)}(x) - \varphi_\mathrm{b}(x) = \sum_{n \geq D+1} c_n \psi_n^{(0)}(x-b) - \varphi_\mathrm{b}(x-b) \,.
\end{equation}
This change of coordinates yields the Jacobian matrix
\begin{equation}
    J = \left( \frac{\partial a_i}{\partial b_k} ~ \Biggr|~ \frac{\partial a_i}{\partial c_j} \right) \,,
\end{equation}
where the $i \geq 1$ are the row indices and $k=1, \ldots, D$ and $j \geq D+1$ are the column indices. Then, following the derivation presented in appendix D of \cite{Andreassen:2016cvx}, we arrive at
\begin{align}
    |\det J| = \left| \det \left( \braket{\psi_i^{(0)} | \partial_k \varphi_\mathrm{b}} + \sum_{j \geq D+1} c_j \braket{\psi_i^{(0)} | \partial_k \psi_j^{(0)}} \right)_{1 \leq i,k \leq D} \right| \,,
\end{align}
where we introduced the bra-ket notation and defined the inner product in the space $L^2(\mathbb{R}^D)$, $\braket{f|g} = \int_x f(x) g(x)\,$.
Notice that the index $i$ runs over the zero modes only ($1\leq i \leq D$), so $\psi_i^{(0)} = \partial_i \varphi_0 / \mathcal{N}$, where $\mathcal{N}$ is the norm of these modes. By Derrick's scaling argument \cite{Derrick:1964ww}, one can show that since $\varphi_0$ is a stationary point of $S_0$, $D S_0[\varphi_0] = \int_x (\partial_i \varphi_0)^2$. Also, given that $\varphi_0$ is radial, there is an equipartition in each of the $D$ terms in the Einstein sum $(\partial_i \varphi_0)^2$, so that $\mathcal{N} = \sqrt{S_0[\varphi_0]}$.

Altogether, the first term in the determinant reads (no summation implied)
\begin{equation}
    \braket{\psi_i^{(0)} | \partial_k \varphi_\mathrm{b}} = \frac{1}{\sqrt{S_0[\varphi_0]}} \int_x \partial_k \varphi_0(x) \partial_i \varphi_\mathrm{b}(x) = \delta_{ik} \frac{1}{\sqrt{S_0[\varphi_0]}} \int_x \partial_i \varphi_0(r) \partial_i \varphi_\mathrm{b}(r) \,.
    \label{eq:cosita}
\end{equation}
The second term in the determinant corresponds to higher-order corrections in the Jacobian, and they can be computed in a diagrammatic way as detailed in \cite{Ekstedt:2022tqk}. Here we omit them, as we aim to obtain the NLO approximation to the statistical rate. Therefore, we are only left with the determinant of \eqref{eq:cosita}. Since this term depends on $\varphi = \varphi_0 + \epsilon \varphi_1$, we must also expand it to $\mathcal{O}(\epsilon)$:
\begin{equation}
    |\det J| \thickapprox \left| \det \left( \braket{\psi_i^{(0)} | \partial_k \varphi_\mathrm{b}} \right)_{1 \leq i,k \leq D} \right| \thickapprox \left( S_0[\varphi_0] \right)^{D/2} \left( 1 + \epsilon\, \frac{1}{S_0[\varphi_0]} \int_x \sum_{i=1}^D \partial_i \varphi_0 \partial_i \varphi_1 \right)\,,
\end{equation}
where we have explicitly written a sum over spatial indices in the remaining integral to avoid confusion. We omit the absolute value in the second factor based on the assumption that the small $\mathcal{O}(\epsilon)$ term cannot possibly change the sign of the $\mathcal{O}(\epsilon^0)$. 

We therefore see that HDOs also enter the zero-mode Jacobian through the perturbative correction to the bounce, $\varphi_1$. We will see in an example in section \ref{sec:ex 1} that this $\mathcal{O}(\epsilon)$ contribution is essential to guarantee the perturbative field redefinition invariance of $\Gamma_\mathrm{stat}$.

Finally, we can write
\begin{align}
    e^{-S_\mathrm{eff}[\varphi_\mathrm{b}]} &\thickapprox \mathcal{V}(\mathbb{R}^D) \left( S_0[\varphi_0] \right)^{D/2} e^{-S_0[\varphi_0]} \int \Dpath \eta'\, e^{-\frac{1}{2} \eta S_0''[\varphi_0]\eta} \times \nonumber\\
    & \hphantom{\thickapprox} \left\{ 1 - \epsilon \left[ \frac{-1}{S_0[\varphi_0]} \int_x \sum_{i=1}^D \partial_i \varphi_0 \partial_i \varphi_1 + S_1[\varphi_0] \right.\right. \nonumber\\
    &\hphantom{\thickapprox \Bigl\{ 1 - \epsilon \Bigl[}\left.\left. + \frac{1}{2} \frac{\int \Dpath \eta'\, e^{-\frac{1}{2} \eta S_0''[\varphi_0]\eta} \eta \left( S_1''[\varphi_0] + S_0'''[\varphi_0] \varphi_1 \right) \eta}{\int \Dpath \eta'\, e^{-\frac{1}{2} \eta S_0''[\varphi_0]\eta}} \right] \right\}\,,
    \label{eq:Seff expanded 2}
\end{align}
where $\Dpath \eta' = \prod_{i \geq D+1} d c_i$ denotes that zero-modes are not integrated over, and the (infinite) volume factor comes from the additional integration over the position of the bounce center $\int d^D b = \mathcal{V}(\mathbb{R}^D)$. Since we are actually interested in the nucleation rate per unit volume, we need not worry about this factor when we take the thermodynamic limit ($\lim_{\mathcal{V} \to \infty} \Gamma/\mathcal{V} < \infty$).

\subsubsection{Functional determinants and traces}
\label{sec:dets and trs}

The zero-mode-subtracted path integrals in eq.~\eqref{eq:Seff expanded 2} can be shown to correspond to functional determinants and traces. To do this, we keep working in a finite volume $\mathcal{V}(\mathbb{R}^D)$, so we can express the fluctuation paths as (infinite-dimensional) linear combinations of the eigenmodes of $S''_0[\varphi_0]$. We have
\begin{align}
    \int \Dpath \eta' \, e^{-\frac{1}{2} \eta S_0''[\varphi_0]\eta} &= \lim_{N \to \infty} \int \prod_{i=D+1}^N d c_i \, e^{-\frac{1}{2} \lambda_i^{(0)} c_i^2} \nonumber\\
    &= \lim_{N \to \infty} \prod_{i=D+1}^N \sqrt{\frac{2 \pi}{\lambda_i^{(0)}}} = \lim_{N \to \infty} \sqrt{\frac{(2 \pi)^{N-D}}{\det' \left( S''_0[\varphi_0] \right)}}\,,
    \label{eq:funcdet}
\end{align}
where we omitted the problematic zero eigenvalues. Notice that the presence of the single negative eigenvalue turns this square root purely imaginary.

Similarly, the ratio of path integrals at $\mathcal{O}(\epsilon)$ in eq.~\eqref{eq:Seff expanded 2} corresponds to a functional trace. Indeed, to show this add a source term to the previous zero-mode-subtracted path integral, where we abbreviate $A \equiv S''_0[\varphi_0]$:
\begin{equation}
    Z(J) \equiv \int \Dpath \eta' e^{-\frac{1}{2} \eta A \eta + J \eta} = \int \Dpath \eta' e^{-\frac{1}{2} (\eta - A^{-1} J) A (\eta - A^{-1} J) + \frac{1}{2} J A^{-1} J} = Z(0)\, e^{\frac{1}{2} J A^{-1} J}\,, 
\end{equation}
where $J \eta \equiv \int_x J(x) \eta(x)$. Differentiating with respect to the $i$ and $j$ components of $J$ in the basis of fluctuation modes yields
\begin{equation}
    \frac{\partial^2 Z(J)}{\partial J_i \partial J_j}\Biggr|_{J=0} = \int \Dpath \eta' \eta_i \eta_j e^{-\frac{1}{2} \eta A \eta} = Z(0) \frac{\partial^2}{\partial J_i \partial J_j} e^{\frac{1}{2} J A^{-1} J} \Biggr|_{J=0} = Z(0) (A^{-1})_{ij}\,,
\end{equation}
so, also abbreviating $B \equiv S_1''[\varphi_0] + S_0'''[\varphi_0] \varphi_1$,
\begin{align}
    \frac{\int \Dpath \eta'\, e^{-\frac{1}{2} \eta A \eta} \eta B \eta}{\int \Dpath \eta'\, e^{-\frac{1}{2} \eta A \eta}} &= \lim_{N \to \infty} \frac{1}{Z(0)} \sum_{i, j \geq D+1}^N B_{ij} \int \Dpath \eta' \eta_i \eta_j e^{-\frac{1}{2} \eta A \eta} = \lim_{N \to \infty} \sum_{i, j \geq D+1}^N B_{ij} (A^{-1})_{ij} \nonumber\\
    &= \lim_{N \to \infty} \sum_{i, j \geq D+1}^N (A^{-1})_{ji} B_{ij} = \lim_{N \to \infty} \tr' \left( A^{-1} B \right)\,,
    \label{eq:functr}
\end{align}
where we have used the symmetry of $A^{-1}$ in the second-to-last equality. Note that the trace is primed, which means that no zero modes are summed over, as required for the well-definiteness of the inverse of $A$. 

At last, we are in a position to write the exact $\mathcal{O}(\epsilon)$ expression for the NLO statistical rate. All that is left is to subtract from eq.~\eqref{eq:Seff expanded 2} the factor $e^{-S_\mathrm{eff}[\varphi_\mathrm{FV}]}$. This will, among others, cancel the divergent $(2\pi)^{N/2}$ prefactor of the functional determinant in eq.~\eqref{eq:funcdet}, leaving an extra $(2 \pi)^{-D/2}$. Therefore:
\begin{empheq}[box=\fbox]{align}
    \frac{\Gamma_\mathrm{stat}}{\mathcal{V}} &\thickapprox \left( \frac{S_0[\varphi_0]}{2 \pi} \right)^{D/2} e^{-S_0[\varphi_0]} \,\left|\frac{\det\left( S''_0[0] \right)}{{\det}'\left( S''_0[\varphi_0] \right)}\right|^{1/2} \times \nonumber\\
    & \hphantom{\thickapprox} \Biggl\{ 1 - \epsilon \Biggl[ \frac{-1}{S_0[\varphi_0]} \int_x \sum_{i=1}^D \partial_i \varphi_0 \partial_i \varphi_1 + S_1[\varphi_0] \nonumber\\
    & \hphantom{\thickapprox \Biggl\{} + \frac{1}{2} \tr' \left( \left( S''_0[\varphi_0] \right)^{-1} \left( S_1''[\varphi_0] + S_0'''[\varphi_0] \varphi_1 \right)\right) - \frac{1}{2} \tr \left(\left( S''_0[0] \right)^{-1} S_1''[0] \right) \Biggr] \Biggr\}\,,  
\end{empheq}
where we have left the $N\to\infty$ limit implicit, meaning that the functional determinants and trace also probe the continuum part of the spectrum of $S''_0[\varphi_0]$ and $S''_0[0]$. Hereinafter we will simply use the symbol $\Gamma_\mathrm{stat}$ to denote the volume-normalized rate.

\section{Evaluation of the functional traces}

In this section we will lay out a series of techniques and results to evaluate the functional traces that appear at $\mathcal{O}(\epsilon)$ in $\Gamma_\mathrm{stat}$. 

We must evaluate the following difference of functional traces
\begin{equation}
    {\tr}'\left( G \mathcal{M}_{\varphi_0} \right) - \tr \left(G_\mathrm{FV} S''_1[0] \right)\,,
    \label{eq:generic trace}
\end{equation}
where $G \equiv (S''_0[\varphi_0])^{-1}$ is the propagator on the $\varphi_0$ background, $G_\mathrm{FV} \equiv (S''_0[0])^{-1}$ is the propagator on the FV, and $\mathcal{M}_{\varphi_0} \equiv  S_1''[\varphi_0] + S_0'''[\varphi_0] \varphi_1$ and $S''_1[0]$ are vertex functions.

The first trace involves an integral of the product of operators on a non-homogeneous background and thus must be carried out numerically. The second involves a homogeneous-background propagator with vertex insertions. Let us address the computation of these traces, noting that we work in DimReg throughout, with $D=3-2\varepsilon$ (not to be confused with the perturbative EFT parameter, $\epsilon$).

The traces will take the general form
\begin{equation}
    \tr' \left( G \mathcal{M} \right) = \int_x \int_y G(x,y) \mathcal{M}(x,y) \,,
\end{equation}
with $G(x,y), \mathcal{M}(x,y)$ the corresponding (zero-mode subtracted) propagator and vertex function kernels in coordinate space. Only after applying all derivative operators in $\mathcal{M}$ to $G$, we take the coincidence limit $y \to x$, imposed by Dirac deltas inside the local kernels, to transform the traces into single integrals in $x$. This is the infinite-dimensional analogue of summing over the diagonal of the operator $G \mathcal{M}$.

\subsection{Propagators}
\label{sec:propagator}

The bounce propagator satisfies
\begin{equation}
    \left[ -\partial^2 + V''_0(\varphi_0) \right] G(x,y) = \delta^{D}(x-y)\,,
    \label{eq:full propagator equation}
\end{equation}
where $\varphi_0 = \varphi_0(r)$ is $\mathcal{O}(D)$-symmetric. We exploit this radial symmetry by decomposing $G(x,y)$ in $D$-dimensional spherical harmonics $Y_{\ell a}(\Omega)$, so that
\begin{equation}
    G(x,y) = \sum_{\ell=0}^\infty \sum_{a=1}^{d_\ell^{(D)}} Y_{\ell a}(\Omega) Y_{\ell a}^*(\Omega') G_\ell(r, r')
    = \frac{1}{\Omega_{D-1}}  \sum_{\ell=0}^\infty  d_\ell^{(D)} \frac{C_\ell^\lambda(\cos\alpha)}{C_\ell^\lambda(1)} G_\ell(r, r') \,,
\end{equation}
where $r=|x|, r'=|y|$, $\Omega_{D-1} = \frac{2 \pi^{D/2}}{\Gamma(D/2)}$, $\cos\alpha = \frac{x\cdot y}{|x||y|}$, $\lambda=(D-2)/2$, $C_\ell^\lambda$ are the Gegenbauer polynomials and $d_\ell^{(D)}$ is the degeneracy factor:
\begin{equation}
    d_\ell^{(D)} = \frac{(2\ell + D -2)(\ell +D -3)!}{\ell! (D-2)!}\,.
    \label{eq:degeneracy factor}
\end{equation}

With this, eq.~\eqref{eq:full propagator equation} becomes a second order ordinary differential equation:
\begin{equation}
    \left[\partial_r^2 + \frac{D-1}{r} \partial_r - \frac{\ell (\ell + D -2)}{r^2} - V_0''(\varphi_0(r)) \right] G_\ell(r, r') = -\frac{1}{r^{D-1}}\delta(r-r')\,.
    \label{eq:spherical harmonic propagator equation}
\end{equation}

\paragraph{Zero modes.} Due to the presence of zero modes of $S''_0[\varphi_0]$, for which this operator is non-invertible, there exists no solution $G(x,y)$ for eq.~\eqref{eq:full propagator equation} with the required boundary conditions (see appendix \ref{app:propagator}) unless those modes are projected out.

The spectral decomposition of $S''_0[\varphi_0]$ involves a discrete and a continuum ($\sigma$) spectrum
\begin{equation}
    S''_0[\varphi_0] = \sum_{n \geq 1} \lambda_n^{(0)} \ket{\psi_n^{(0)}} \bra{\psi_n^{(0)}} + \sum_{a} \int_{\sigma}d\lambda^{(0)} \lambda^{(0)} \ket{\psi_{\lambda a}^{(0)}} \bra{\psi_{\lambda a}^{(0)}}\,,
\end{equation}
where $\sum_a$ sums over degeneracies. Its inverse $\widetilde{G}$ exists only on the eigenspace orthogonal to the zero modes, and thus reads in space coordinates
\begin{equation}
    \widetilde{G}(x,y) = \sum_{n \geq D+1} \frac{\psi_n^{(0)}(x) \psi_n^{(0)}(y)}{\lambda_n^{(0)}} + \sum_{a} \int d\lambda^{(0)} \frac{\psi_{\lambda a}^{(0)}(x) \psi_{\lambda a}^{(0)}(y)}{\lambda^{(0)}}\,,
\end{equation}
where the $D$ first eigenmodes $\psi_n^{(0)}$ have zero eigenvalue. Applying $S''_0[\varphi_0]$ to $\widetilde{G}$ therefore defines the equation
\begin{equation}
    \left[ -\partial^2 + V''_0(\varphi_0) \right] \widetilde{G}(x,y) = \delta^{D}(x-y) - \sum_{i=1}^D z_i(x) z_i(y) \equiv \delta^{D}(x-y) - \Pi_0(x,y)\,,
    \label{eq:full propagator tilde equation}
\end{equation}
where $\Pi_0(x,y)$ is the kernel of the zero-mode projector.

The $D$ zero modes $z_i = (S_0[\varphi_0])^{-1/2} \partial_i \varphi_0(r)$ lie in the $\ell=1$ sector, since
\begin{align}
    &\left[-\partial_r^2 - \frac{D-1}{r} \partial_r \right] \varphi_0 + V'_0(\varphi_0) = 0 \nonumber\\
    &\implies\, \left[-\partial_r^2 - \frac{D-1}{r} \partial_r + \frac{D-1}{r^2} + V''_0(\varphi_0) \right] \dot\varphi_0 = 0\,,
\end{align}
that is, $S''_0[\varphi_0] z_i = 0$ for $\ell = 1$ (see eq.~\eqref{eq:spherical harmonic propagator equation}). Therefore, expanding eq.~\eqref{eq:full propagator tilde equation} in spherical harmonics, we finally have that $\widetilde{G}_\ell(r,r')$ satisfies the shifted equation
\begin{multline}
    \left[-\partial_r^2 - \frac{D-1}{r} \partial_r + \frac{\ell( \ell + D - 2)}{r^2} + V_0''(\varphi_0(r)) \right] \widetilde{G}_\ell(r, r') 
    \\
    = \frac{1}{r^{D-1}}\delta(r-r') - \frac{\Omega_{D-1}}{D  S_0[\varphi_0]} \delta_{\ell 1} \dot\varphi_0(r) \dot\varphi_0(r')\,.
    \label{eq:spherical harmonic propagator tilde equation}
\end{multline}
Thus, zero modes are effectively removed from the functional trace by changing $G$ for $\widetilde G$. In appendix \ref{app:propagator} we explain how we solve this differential equation for the propagator in practice.

Finally, note that the FV propagator also satisfies eq.~\eqref{eq:full propagator equation} where $V_0''(\varphi_0) \to V_0''(0) \equiv \omega^2$ is a constant. In this case, there is an analytical form for each partial wave
\begin{equation}
    G_{\mathrm{FV},\ell}(r) = \frac1{r^{D-2}} I_{\ell + \frac{D-2}{2}}(\omega r)\, K_{\ell + \frac{D-2}{2}}(\omega r) \,,
    \label{eq:FV prop}
\end{equation}
where $I_\nu(z)$ and $K_\nu(z)$ are the modified Bessel functions of first and second kind, respectively.

\subsection{Vertex function and general form of the trace}

The vertex function $\mathcal{M}_{\varphi_0}$ will generally not be only a polynomial in $\varphi_{0,1}$ and their derivatives, but also a combination of derivatives acting on delta functions. Upon integration by parts (IBP), these derivatives will turn into derivatives acting on $\widetilde{G}$ or $\varphi_{0,1}$. 

%Alternatively, one can consider $\mathcal{M}_{\varphi_0}$ a differential operator with ``open'' derivatives.

To handle these terms, we make use of eq.~\eqref{eq:full propagator tilde equation} to write the recursive formula
\begin{equation}
    \partial^{2k} \widetilde G(x,y) = \partial^{2(k-1)} \biggl[ V_0''(\varphi_0)  \widetilde G(x,y) + \Pi_0(x,y) - \delta^{D}(x-y) \biggr] \,.
    \label{eq:remove second derivatives over G}
\end{equation}

On the other hand, when a single derivative is acting on $G(x,y)$, we can use the symmetry of the propagator, \textit{i.e.} $\widetilde G(x,y) = \widetilde G(y,x)$, so that after taking the $x \to y$ limit, a derivative acting over one of its arguments can be traded off by half of the derivative of the diagonal:
\begin{equation}
   \left. \partial_{x_i} \widetilde{G}(x,y) \right|_{y=x} = \frac{1}{2} \partial_i \widetilde{G}(x)\,.
   \label{eq:G symmetry}
\end{equation}

In light of the above, the trace is always reduced to an expression containing only factors of $G(x)$ and $\partial_i G(x)$~\footnote{To $\mathcal{O}(\epsilon)$ in the EFT expansion, one can further use IBP identities to remove any derivatives over $G(x)$.}. After expanding in spherical harmonics we will generically deal with
\begin{gather}
    {\tr}'\left( G \, \mathcal{M}_{\varphi_0} \right) = \sum_{\ell=0}^\infty \widetilde{T}_{\mathrm{b}, \ell} \,, \nonumber\\
    \widetilde{T}_{\mathrm{b},\ell} \equiv d_\ell^{(D)} \int_0^\infty dr \, r^{D-1} \left[ \widetilde{G}_\ell (r) \mathcal{A}(r) + \dot{\widetilde{G}}_\ell (r) \mathcal{B}(r) + \delta_{\ell 1} \mathcal{C}(r) + \mathcal{D}(r) \delta(r-r) \right]\,,
    \label{eq:trace general bounce}
\end{gather}
where $\mathcal{A},\mathcal{B},\mathcal{C},\mathcal{D}$ are functions of $r, \varphi_0(r)$ and $\varphi_1(r)$ that contain no free derivatives, and $\delta(r-r)$ is to be understood as $\lim_{r' \to r} \delta(r-r')$. Also, $\dot{\widetilde{G}}_\ell(r) \equiv \partial_r \widetilde{G}_\ell(r)$.

Among these functions, $\mathcal{C}$ and $\mathcal{D}$ can be treated separately. The term $\mathcal{C}(r)$ comes from the second term on the r.h.s. of eq.~\eqref{eq:remove second derivatives over G}, so it only appears at $\ell = 1$. On the other hand, $\mathcal{D}(r)$ contains terms that stem from removing derivatives from the $\delta(r-r)$ (\textit{e.g.}, from $k=2$ in eq.~\eqref{eq:remove second derivatives over G}) upon IBP. The product $\mathcal{D}(r) \delta(r-r)$ is divergent, but vanishes in DimReg with the usual prescription (see end of appendix~\ref{app:fr invariance}).

As shown in appendix \ref{app:radial integrals}, the radial integrals at each fixed $\ell$ are divergent unless one subtracts the corresponding integrals from the FV trace. We then work out their difference instead of evaluating them separately in what follows. Note that the FV trace is considerably simpler. In particular, the trace will always boil down to
\begin{gather}
    \tr \left( G_\mathrm{FV} S''_1[0] \right) = \sum_{\ell=0}^{\infty} T_{\mathrm{FV}, \ell} \,, \nonumber\\
    T_{\mathrm{FV}, \ell} \equiv d_\ell^{(D)} \int_0^\infty dr \, r^{D-1} \left[ \mathcal{A}_{\rm FV} G_{\mathrm{FV}, \ell} (r)  + \mathcal{D}_{\rm FV} \delta(r-r) \right]\,.
    \label{eq:trace general fv}
\end{gather}
The second term will cancel in DimReg as discussed above. Meanwhile, the constant $\mathcal{A}_{\rm FV}$ comprises the potential piece $V_1''(0)$, as well as terms coming from the removal of an even number of derivatives acting on $G_{\mathrm{FV}, \ell} (r)$, by means of eq.~\eqref{eq:remove second derivatives over G}. In particular, a term $\partial^{2k}$ inside $S_1''[0]$ will amount to a multiplicative factor $V_0''(0)^k$. Notice that a contribution of the kind $\mathcal{B}_{\rm FV}(r) \dot G_{\mathrm{FV}, \ell} (r) $ is not possible, since an odd number of derivatives over the propagator has to compensate with at least one derivative over $\varphi$, which vanishes for $\varphi=\varphi_\mathrm{FV}$.

\subsection{Tackling the infinite sum}
\label{subsec:tackling the infinite sum}

The difference between the bounce and FV traces can then be expressed as
\begin{equation}
    {\tr}'\left( G \mathcal{M}_{\varphi_0} \right) - \tr \left(G_\mathrm{FV} S''_1[0] \right) = \sum_{\ell=0}^\infty \widetilde{T}_\ell \equiv \mathcal{S} \,,
\end{equation}
where $\widetilde{T}_\ell \equiv \widetilde{T}_{\mathrm{b},\ell} - T_{\mathrm{FV},\ell}$. In the large-$\ell$ limit we can approximate $\widetilde{G}_\ell(r)$ using the WKB method (see appendix \ref{app:propagator}), but this does not avoid the divergence at large $\ell$. We will therefore have to regulate the infinite sum $\mathcal{S}$.

Before that, we follow \cite{Ekstedt:2022tqk} to add and subtract the WKB-approximated integrals so that
\begin{equation}
    \mathcal{S} = \sum_{\ell=0}^\infty \left(\widetilde{T}_\ell - T_{\ell}^{\rm (WKB)} \right) + \sum_{\ell=0}^\infty T_{\ell}^{\rm (WKB)} \,.
    \label{eq:splitting of sum (pre)}
\end{equation}
Here $T_{\ell}^{\rm (WKB)}$ shares the same definition as $\widetilde{T}_{\ell}$, with the exception that every $\widetilde{G}_\ell(r)$ is to be replaced with the WKB-approximated solution to the corresponding propagator equation, $G_\ell^{\rm (WKB)}(r)$. This propagator takes the following form for both the bounce and the FV backgrounds, after inserting the corresponding $\varphi_0$ (see appendix~\ref{app:propagator} for details):
\begin{equation}
    G_\ell^{\rm (WKB)}(r) = \frac{1}{D+2\ell - 2} r^{2-D} - \frac{2}{(D+2\ell - 2)^3} r^{4-D} V_0''(\varphi_0)\,.
    \label{eq:WKB approximation propagator}
\end{equation}
Since $T_{\ell}^{\rm (WKB)} \sim \widetilde{T}_{\ell}$ in the large-$\ell$ limit, we effectively sum their difference up to a fixed $L\gg 1$. We then approximate the trace difference as
\begin{equation}
    \mathcal{S} \approx \sum_{\ell=0}^L \left(\widetilde{T}_\ell - T_{\ell}^{\rm (WKB)} \right) + \sum_{\ell=0}^\infty T_{\ell}^{\rm (WKB)} \,.
    \label{eq:splitting of sum (post)}
\end{equation}

The finite sum is computed by numerically solving the corresponding differential equation for $\widetilde{T}_\ell$ and performing the numerical integration for each $\ell \leq L$, as there is no divergence in this term. Due to the infinite series in $\ell$, the second term does need to be regularized. Using eqs.~\eqref{eq:degeneracy factor}, \eqref{eq:trace general bounce}, \eqref{eq:trace general fv}, \eqref{eq:WKB approximation propagator}, this sum will be of the form
\begin{equation}
    \sum_{\ell=0}^\infty T_{\ell}^{\rm (WKB)} = \frac{1}{(D-2)!}
    \int_0^\infty dr\, r^{D-1} \sum_{\ell=0}^\infty \left[ \frac{(\ell + D - 3)!}{\ell!} \Psi(r) + \frac{(\ell + D - 3)!}{\ell!(D + 2\ell - 2)^2} \Omega(r)  \right]\,,
\end{equation}
where the two terms inside the sum come from each of the terms in eq.~\eqref{eq:WKB approximation propagator}, and they read
\begin{align}
    \Psi(r) &= r^{2-D} \left(\mathcal{A}(r) -\mathcal{A}_{\rm FV}\right) + (2-D) \mathcal{B}(r) r^{1-D}\,,
    \\
    \Omega(r) &= - 2 r^{4-D} \left\{ \left[\mathcal{A}(r) V_0''(\varphi_0) - \mathcal{A}_{\rm FV} V_0''(0) \right] + \mathcal{B}(r) \left[ \frac{4-D}{r} V_0''(\varphi_0)  + V_0'''(\varphi_0) \dot \varphi_0(r) \right] \right\} \,.
\end{align}
Notice that there is no need to include the $\mathcal{C}(r)$ factor present only when $\ell=1$ in the definition of $T_\ell^{\rm (WKB)}$, since it will cancel with the same term in the finite sum in eq.~\eqref{eq:splitting of sum (post)} as long as $L\geq1$.

Let us consider each of the two terms in the sum separately.

\paragraph{First sum.} This sum is divergent for $D>2$, so we introduce a regulator $x^\ell$ so that
\begin{equation}
 \left[ \sum_{\ell=0}^\infty \frac{(\ell + D - 3)!}{\ell!} \right]_{\rm DimReg} = \lim_{x\to1^{-}} \sum_{\ell=0}^\infty \frac{(\ell + D - 3)!}{\ell!} x^\ell = \frac{(D-3)!}{(1-x)^{D-2}}\,.
    \label{eq:regulated first sum}
\end{equation}
For $D < 2$, the r.h.s. vanishes in the $x\to 1^{-}$ limit. Therefore, we perform an analytic continuation so that the sum is vanishing in DimReg for $D\geq2$ too.

\paragraph{Second sum.} This sum is convergent in $D=3$. Precisely, 
\begin{equation}
    \sum_{\ell=0}^\infty \frac{(\ell + D - 3)!}{\ell!(D + 2\ell - 2)^2} \Biggr|_{D=3} = \frac{\pi^2}{8} \,.
\end{equation}

For completeness, we give the dimensionally-regularized value of the infinite sum for $D>3$. By means of the identity
\begin{equation}
    \frac{1}{(D+2\ell-2)^2} = -\int_0^1 dt \, t^{D + 2\ell -3} \log t \,,
\end{equation}
we write
\begin{align}
\notag
    \left[ \sum_{\ell=0}^\infty \frac{(\ell+D-3)!}{\ell! (D+2\ell-2)^2 } \right]_{\rm DimReg} &= - \int_0^1 dt \, t^{D-3} \log t \sum_{\ell=0}^\infty \frac{(\ell+D-3)!}{\ell!} t^{2\ell} 
    \\ \notag
    & = - (D-3)! \int_0^1 dt \, t^{D-3} (1-t^2)^{2-D} \log t
   \\ & = 2^{-D} \sqrt{\pi} \cot{\left(\frac{\pi D}{2}\right)} \Gamma(D-2) \Gamma\left(\frac{D-2}{2}\right) \Gamma\left(\frac{3-D}{2}\right) \,,
\end{align} 
which can be assigned a value for all $D$.

All in all, in $D=3$ we have:
\begin{equation}
    \sum_{\ell=0}^\infty T_{\ell}^{\rm (WKB)} = -\frac{\pi^2}{4}
    \int_0^\infty dr\, r^3 \left[ \mathcal{A}(r) V_0''(\varphi_0) - \mathcal{A}_{\rm FV} V_0''(0) + \mathcal{B}(r) \left(\frac{V_0''(\varphi_0)}{r}  + V_0'''(\varphi_0) \dot \varphi_0(r)  \right)\right],
\end{equation}
and the concrete form of $\mathcal{A}(r), \mathcal{A}_{\rm FV}$ and $\mathcal{B}(r)$ is model-dependent.

\section{Applications}
\label{sec:ex 1}

\subsection{Example 1: A higher-dimensional operator with derivatives}

In this example we address the explicit computation of the statistical rate through $\mathcal{O}(\epsilon)$ in a model containing a derivative operator different from the kinetic term, and its invariance under field redefinitions.

\subsubsection{Setup}

Consider the following 3D EFT action
\begin{equation}
    S[\phi] = S_0[\phi] + S_1[\phi] \,,
\end{equation}
where $S_0[\phi] = \int_x \left[ \frac{1}{2} (\partial_i \phi)^2 + V_0(\phi) \right]$ and
\begin{align}
    V_0(\phi) &= \frac{1}{2} m_3^2 \phi^2 + \frac{\kappa_3}{3!} \phi^3 + \frac{\lambda_3}{4!} \phi^4 \,, \\
    S_1[\phi] &= \frac{\alpha}{2} \int_x \phi^2 (\partial_i \phi)^2 \,,
\end{align}
with a perturbative $\alpha$ that we keep arbitrary. For concreteness, we fix the following values for the 3D parameters, in units of mass $m_3$:
\begin{equation}
    m_3^2 = 1\,,\quad \kappa_3 = -2\,,\quad \lambda_3 = \frac13\,.
    \label{eq:ex1 params}
\end{equation}

\subsubsection{Numerical evaluation}

First, we compute the $\mathcal{O}(\epsilon^0)$ bounce solution $\varphi_0$ using \texttt{CosmoTransitions} \cite{Wainwright:2011kj} together with its action $S_0[\varphi_0]$. Then, we use \texttt{BubbleDet} \cite{Ekstedt:2023sqc} to compute the $\mathcal{O}(\epsilon^0)$ functional determinant. What is left is to compute the $\mathcal{O}(\epsilon)$ contributions to $\Gamma_\mathrm{stat}$ (we will drop the subscript ``stat'' hereinafter). 

Since $\varphi_{\mathrm{FV},i} = 0$ for $i=0,1$, the $\mathcal{O}(\epsilon)$ FV contribution is vanishing, and
\begin{align}
    \log \Gamma &= \log \Gamma^{(0)} + \log \Gamma^{(1)} \nonumber\\
    &= \log \Gamma^{(0)} - \left[ A - j + \frac{1}{2}(\mathcal{T}_1 + \mathcal{T}_2) \right]  \,,
\end{align}
where $\Gamma^{(0)}$ is the $\mathcal{O}(\epsilon^0)$ statistical rate and
\begin{align}
    A &= S_1[\varphi_0] = 2\pi \alpha \int_0^\infty dr\, r^2 \varphi_0(r)^2 \dot \varphi_0(r)^2 \,, \\
    j &= \frac{4 \pi \alpha}{S_0[\varphi_0]} \int_0^\infty dr \, r^2 \dot \varphi_0(r) \dot \varphi_1(r) \,, \\
    \mathcal{T}_1 &= \tr' \left( \widetilde{G} S_1''[\varphi_0] \right) \,, \\
    \mathcal{T}_2 &= \tr' \left( \widetilde{G} V_0'''(\varphi_0) \alpha \varphi_1 \right) \,.
\end{align}
Here, $\varphi_0, \varphi_1$ satisfy the radial differential equations
\begin{align}
    \ddot \varphi_0(r) + \frac{2}{r} \dot \varphi_0(r) - V_0'(\varphi_0(r)) &= 0 \,, 
    \label{eq:phi0}
    \\
    \ddot \varphi_1(r) + \frac{2}{r} \dot \varphi_1(r) - V_0''(\varphi_0(r)) \varphi_1(r) &= \frac{1}{\alpha} S_1'[\varphi_0(r)] \,,
    \label{eq:phi1}
\end{align}
with boundary conditions
\begin{equation}
    \dot \varphi_0(0) = \dot \varphi_1(0) = \varphi_0(\infty) = \varphi_1(\infty) = 0 \,.
\end{equation}
We explicitly removed the $\alpha$-dependence from $\varphi_1$, so that $\varphi = \varphi_0 + \alpha \varphi_1$. On the other hand, $\widetilde{G}(x,y)$ obeys eq.~\eqref{eq:full propagator tilde equation} (see appendix \ref{app:propagator}).

For the benchmark \eqref{eq:ex1 params} we obtain
\begin{equation}
    S_0[\varphi_0] = 55.95068(5)
    \,,
    \qquad
    \log\Gamma^{(0)} = -61.7815(5)\,,
\end{equation}
where we indicate the numerical error on the last significant digit. From $\varphi_0$, the evaluation of $A$ is also straightforward:
\begin{equation}
    \frac{A}{\alpha} = 292.4793(1)\,.
\end{equation}

Next, we turn to the computation of $\varphi_1$. The first derivative of $S_1$ reads
\begin{align}
\notag
    \frac{1}{\alpha} S_1'[\varphi_0] &= - \varphi_0(r) \dot \varphi_0(r)^2 -  \varphi_0(r)^2 \left(\ddot \varphi_0(r) + \frac{2}{r}\dot \varphi_0(r) \right) \\
    &= - \varphi_0(r) \dot \varphi_0(r)^2 - \varphi_0(r)^2 V_0'(\varphi_0(r)) \,,
\end{align}
where we used the EOM of $\varphi_0(r)$, thus minimizing errors from the numerical differentiation of $\varphi_0$. Eq.~\eqref{eq:phi1} can be solved with any standard numerical method. Once $\varphi_1(r)$ is found, the numerical integration of $j$ is straightforward. We find
\begin{equation}
     \frac{j}{\alpha} = 2.61373(1) \,.
\end{equation}

In order to evaluate $\mathcal{T}_1$, we deal with
\begin{multline}
    \frac{1}{\alpha} S_1''[\varphi_0](x,y) = 
    - \left[(\partial_i\varphi_0(x))^2 + 2 \varphi_0(x) \partial^2 \varphi_0(x) \right] \delta(x-y) \\
    - 2 \varphi_0(x) \partial_i \varphi_0(x) \partial_i \delta(x-y) - \varphi_0(x)^2 \partial^2 \delta(x-y) \,,
\end{multline}
where all partial derivatives act on $x$. Upon IBP, we end up with
\begin{multline}
    \mathcal{T}_1 = \alpha \int_x \int_y \delta(x-y) \bigg\{ -  \left[(\partial_i\varphi_0(x))^2 + 2 \varphi_0(x) \partial^2 \varphi_0(x) \right] \widetilde G(x,y) 
    \\ - 2 \varphi_0(x) \partial_i \varphi_0(x) \partial_i \widetilde G(x,y) - \varphi_0(x)^2 \partial^2 \widetilde G(x,y) \bigg\}\,.
\end{multline}
Using eqs.~\eqref{eq:remove second derivatives over G} and \eqref{eq:G symmetry}, and then taking the coincidence limit we find
\begin{multline}
    \mathcal{T}_1 = \alpha \int_x  \bigg\{ -  \left[(\partial_i\varphi_0(x))^2 + 2 \varphi_0(x) \partial^2 \varphi_0(x) + \varphi_0(x)^2 V_0''(\varphi_0(x)) \right] \widetilde G(x) 
    \\ - \varphi_0(x) \partial_i \varphi_0(x) \partial_i \widetilde G(x) + \varphi_0(x)^2 \delta(x-x) - \varphi_0(x)^2 \Pi_0(x)  \bigg\}\,,
\end{multline}
where $\Pi_0(x)$ is the coincident limit of the zero-mode projector defined in eq.~\eqref{eq:full propagator tilde equation}. The term involving $\delta(x-x)$ vanishes in DimReg. Also, if we apply IBP to remove the derivative over $\widetilde{G}(x)$ and use the EOM for $\varphi_0$, the expression above simplifies to
\begin{equation}
    \mathcal{T}_1 = - \alpha \int_x  \bigg\{\left[\varphi_0(x) V_0''(\varphi_0(x)) + V_0'(\varphi_0(x))\right]\varphi_0(x) \widetilde G(x)
+\varphi_0(x)^2\Pi_0(x) \bigg\} \,.
\end{equation}

Expanding now in partial waves, 
\begin{align}
\notag
    \mathcal{T}_1 = & - \frac{4\pi \alpha}{S_0[\varphi_0]} \int_0^\infty dr\, r^2 \varphi_0(r)^2 \dot \varphi_0(r)^2  
    \\ &~ - \alpha \sum_{\ell=0}^\infty (2\ell + 1) \int_0^\infty dr\, r^2  \left[V_0'(\varphi_0(r)) + \varphi_0(r) V_0''(\varphi_0(r))  \right] \varphi_0(r) \widetilde G_\ell(r) \,,
\end{align}
where $\widetilde G_\ell(r)$ satisfies eq.~\eqref{eq:spherical harmonic propagator tilde equation}. Following the derivation in section~\ref{subsec:tackling the infinite sum}, we approximate $\mathcal{T}_1$ as
\begin{equation}
    \mathcal{T}_1 \thickapprox \sum_{\ell=0}^L T^{(\Delta)}_\ell + \sum_{\ell=0}^\infty T_\ell^{\rm (WKB)} - \frac{4\pi \alpha}{S_0[\varphi_0]} \int_0^\infty dr\, r^2 \varphi_0(r)^2 \dot \varphi_0(r)^2\,,
    \label{eq:split of T1 example}
\end{equation}
where
\begin{equation}
    T_\ell^{(X)} = - \alpha (2\ell + 1) \int_0^\infty dr\, r^2  \left[V_0'(\varphi_0(r)) + \varphi_0(r) V_0''(\varphi_0(r))  \right] \varphi_0(r) G^{(X)}_\ell(r) \,,
    \label{eq:equation for Tl example}
\end{equation}
and $G_\ell^{(\Delta)} = \widetilde G_\ell^{(N)} - G_\ell^{\rm (WKB)}$ with $\widetilde G_\ell^{(N)}$ being the numerical solution to eq.~\eqref{eq:spherical harmonic propagator tilde equation} and $G_\ell^{\rm (WKB)}$ the WKB-approximated solution, namely eq.~\eqref{eq:WKB approximation}.

The second, infinite sum needs to be computed in DimReg. We set $D=3$ in the expression above for simplicity, but we follow the discussion of section~\ref{subsec:tackling the infinite sum} with generic $D$. As shown in said section, the infinite sum reads
\begin{align}
    \frac{1}{\alpha}\sum_{\ell=0}^\infty T_\ell^{\rm (WKB)} &=   \frac{\pi^2}{4}\int_0^\infty dr\,
    r^3 \varphi_0(r)^2 \left(2 m_3^2  + \frac{3}{2} \kappa_3 \varphi_0(r)  + \frac{2}{3}\lambda_3 \varphi_0(r)^2\right)
    \nonumber\\
    & \hphantom{ - \frac{\pi^2}{4}\int_0^\infty dr\,
    r^3 \varphi_0(r)^22} \times \left( m_3^2 + \kappa_3 \varphi_0(r) +  \frac{\lambda_3}{2} \varphi_0(r)^2 \right)  
    \nonumber\\
    &= 401.2730(1) \,.
\end{align}
For the numerical sum, we integrate $T_\ell^{(\Delta)}$ numerically and fix $L=100$. We finally obtain for each term in eq.~\eqref{eq:split of T1 example}, combining the errors in quadrature:
\begin{equation}
    \frac{1}{\alpha}\mathcal{T}_1 = -397.9018(2) + 401.2730(1) - 10.454894(4) = -7.0836(3) \,.
\end{equation}

We repeat this procedure for $\mathcal{T}_2$, which is simpler as $V_0'''[\varphi_0]$ contains no derivatives. We obtain in this case
\begin{align}
    \frac{1}{\alpha} \mathcal{T}_2 &=
    \begin{multlined}[t]
         \sum_{\ell=0}^L (2\ell+1) \int_0^\infty dr\, r^2 \left[\widetilde{G}_\ell^{(N)}(r) - G_\ell^{\rm (WKB)}(r) \right] V_0'''(\varphi_0(r)) \varphi_1(r) 
        \\
        - \frac{\pi^2}{4} \int_0^\infty dr\, r^3 V_0''(\varphi_0(r)) V_0'''(\varphi_0(r)) \varphi_1(r) 
    \end{multlined} \nonumber
    \\
    &= -473.281(1) + 617.1842(3) = 143.904(1) \,.
\end{align}

% \begin{table}[h]
%     \centering
%     \begin{tabularx}{0.8\textwidth}{YYYY}
%     \toprule
%         $A$ & $-j$ & $\frac{1}{2}\mathcal{T}_1$ & $\frac{1}{2}\mathcal{T}_2$  \\
%         \midrule
%         427.9817(2) & -3.19801(2) & -3.7354(2) & 86.4708(8) \\ \bottomrule
%     \end{tabularx}
%     \caption{\small Summary of quantities contributing to the NLO part in the EFT expansion. All quantities are to be understood as computed per unit of $\alpha$.}
%     \label{tab:placeholder}
% \end{table}%

Adding up every contribution we finally obtain
\begin{equation}
    \log\Gamma = -61.7815(5) - 358.2756(8) \, \alpha \,.
    \label{eq:final loggamma}
\end{equation}
The quoted uncertainties throughout are numerical estimates arising from solver tolerances, quadratures and partial-wave truncations, as well as errors from \texttt{BubbleDet}.

\begin{table}[h]
    \centering
    \begin{tabularx}{0.95\textwidth}{YYYY}
    \toprule
        LO--$\epsilon^0$ & NLO--$\epsilon^0$ & LO--$\epsilon$ & NLO--$\epsilon$  \\
        \midrule
        $-55.9507$ & $-5.8309$ & $-292.4793\,\alpha$ & $-65.7963\,\alpha$ \\ \bottomrule
    \end{tabularx}
    \caption{\small Values of the different contributions to $\log\Gamma$ for benchmark \eqref{eq:ex1 params}. Results are tagged as $\mathrm{N}^m\mathrm{LO}$--$\epsilon^n$ referring to the order in the loop and the 3D EFT expansion, respectively.}
    \label{tab:ex1 comparison}
\end{table}

Note that the ratio between NLO and LO is $\mathcal{O}(10\%)$ at $\mathcal{O}(\epsilon^0)$, and it increases up to $\mathcal{O}(25\%)$ at $\mathcal{O}(\epsilon)$. Therefore, if HDOs become relevant, their one-loop contribution to the nucleation rate should be included along with the correction $S_1[\varphi_0]$ to the exponential factor. 

While the individual contributions at $\mathcal{O}(\epsilon)$ might seem large, they are multiplied by a small parameter $\alpha$. The validity of the 3D EFT will depend solely on the smallness of this parameter: unless $\alpha \lesssim 0.1$, the HDO correction dominates over the superrenormalizable term, making the EFT expansion unreliable.

\subsubsection{Field redefinition invariance}

We can explicitly check the FR invariance of this result by introducing the $\mathcal{O}(\epsilon^0)$ EOM for $\phi$ to remove the redundant higher-derivative operator in favor of the physical operators $\hat c_{\phi^5} \phi^5$, $\hat c_{\phi^6} \phi^6$ and corrections to the superrenormalizable terms. The modified WCs read, to $\mathcal{O}(\epsilon)$ \cite{Chala:2024xll}:
\begin{align}
    \hat \lambda_3 &= \lambda_3 - 4 \alpha m_3^2 \,, \label{eq:frd wcs1}\\
    \hat c_{\phi^5} &=  -\frac{\alpha}{12} \kappa_3 \,, \\
    \hat c_{\phi^6} &= -\frac{\alpha}{36} \lambda_3\,, \label{eq:frd wcs3}
\end{align}
while $\hat m_3^2 = m_3^2$ and $\hat \kappa_3 = \kappa_3$. Introducing now all terms proportional to $\alpha$ in the $\mathcal{O}(\epsilon)$ correction to $\Gamma$, we obtain the same result as in tab.~\ref{tab:ex1 comparison}. This is a result of a delicate cancellation of the variation in the $\mathcal{O}(\epsilon)$ piece of the zero-mode Jacobian $j$ and the same variation in the zero-mode-subtracted functional traces $\mathcal{T}_{1,2}$. The statistical rate is therefore exactly FR-invariant up to the order of truncation.

\subsection{Example 2: Real scalar-Yukawa model}

\subsubsection{Setup}

We now consider a 4D model with a real scalar $\phi$ and a Dirac fermion $\psi$,
\begin{equation}
    \mathcal{L}_\mathrm{4D} = \frac{1}{2} (\partial_\mu \phi)^2 - \frac{1}{2} m^2 \phi^2 - \frac{\kappa}{3!} \phi^3 - \frac{\lambda}{4!} \phi^4 + \overline{\psi} i \slashed{\partial} \psi - y \phi \overline{\psi} \psi\,.
    \label{eq:yukawa model}
\end{equation}
This model was originally studied in \cite{Gould:2023jbz} in the context of PTs, and its DR up to dimension-8 (corresponding to $\mathcal{O}(\epsilon^2)$ in our counting in HDOs) was carried out in \cite{Chala:2024xll,Chala:2025oul}. These latter works proved that HDOs significantly affect the critical thermodynamics of PTs in the moderately-strong transition regime, leading to the eventual breakdown of the high-$T$ approximation.

Through $\mathcal{O}(\epsilon)$, the high-$T$ limit of the 4D theory \eqref{eq:yukawa model} is given by a scalar 3D EFT action\footnote{With a slight abuse of notation, we use the same symbol $\phi$ for the 4D scalar and its 3D zero Matsubara mode in the 3D EFT.} $S[\phi] = S_0[\phi] + S_1[\phi]$, where $S_0[\phi] = \int_x \left[ \frac{1}{2} (\partial_i \phi)^2 + V_0(\phi) \right]$ and
\begin{align}
    V_0(\phi) &= \sigma_3 \phi + \frac{1}{2} m_3^2 \phi^2 + \frac{\kappa_3}{3!} \phi^3 + \frac{\lambda_3}{4!} \phi^4 \,, \\
    S_1[\phi] &= \int_x \left[ c_{\phi^5} \phi^5 + r_{\partial^2 \phi^3} \phi^2 \partial^2 \phi +  c_{\phi^6} \phi^6 + r_{\partial^4 \phi^2} \partial^2 \phi \partial^2 \phi + r_{\partial^2 \phi^4} \phi^3 \partial^2 \phi \right] \,,
\end{align}
where we treat both dimension-5 and dimension-6 operators as $\mathcal{O}(\epsilon)$ because they are generated at the same order in our power counting in DR, as we show below. The operators labeled as $c_X$ are physical while the $r_X$ are redundant, in the sense that the latter can field-redefined into the former.

We assume the same benchmark point for the 4D theory parameters used in \cite{Navarrete:2025yxy}:
\begin{align}
    m^2 = 20000 \,\mathrm{GeV}^2\,,\quad \kappa = -240 \,\mathrm{GeV}\,,\quad \lambda = 0.24\,,
    \label{eq:ex2 params}
\end{align}
and, for completeness, we reproduce here their matching equations\footnote{We assume a power counting defined by $m^2/T^2 \sim p^2/T^2 \sim \lambda \sim \kappa/T \sim y^2$. We have fixed typos in $\sigma_3$ and $m_3^2$ in \cite{Navarrete:2025yxy}, which made the WCs remain matching scale dependent after introducing the running of 4D parameters.} and take the leading dimension-5 and -6 matching from \cite{Chala:2024xll}, that we crosscheck with \texttt{Matchotter} \cite{Fuentes-Martin:2026bhr}:
\begin{align}
    \sigma_3 &=\frac{T^{3/2}\kappa}{24}
    +\frac{\kappa}{(4\pi)^2 \sqrt{T}}\Bigg[
    -\frac{1}{2}L_b m^2
    +T^2\left(
    -\frac{L_b\lambda}{16}
    +\left(\frac{1}{2}\ln 2-\frac{L_f}{8}\right)y^2
    \right)
    \nonumber\\
    & -\frac{\lambda T^2}{6}
    \left(c+\ln\frac{3T}{\overline\Lambda_{\mathrm{3d}}}\right)
    \Bigg]\,,
    \\
    m_3^2
    &= m^2+\frac{T^2}{24}\left(\lambda+4y^2\right)
    +\frac{T^2}{(4\pi)^2}\Bigg[
    -\frac{1}{2}L_b \frac{\kappa^2}{T^2}
    -\frac{1}{2}\lambda \frac{m^2}{T^2} L_b
    -2L_f y^2 \frac{m^2}{T^2}
    \nonumber\\
    &
    -\frac{L_b\lambda^2}{16}
    +\left(\frac{L_f}{3}-\frac{4}{3}\ln 2\right)y^4
    +\left(\frac{\ln 2}{2}-\frac{L_f}{6}\right)y^2\lambda
    -\frac{\lambda^2}{6}
     \left(c+\log\frac{3T}{\overline\Lambda_{\mathrm{3d}}}\right)
    \Bigg]\,,
    \\
    \kappa_3    &=
    \kappa \sqrt{T} +\frac{\sqrt{T}}{(4\pi)^2}
    \left(-\frac{3}{2}\kappa\lambda L_b
          -3\kappa y^2L_f\right) \,,
    \\
    \lambda_3
    &=
    \lambda T+\frac{T}{(4\pi)^2}
    \left(-\frac{3}{2}\lambda^2L_b
          +24y^4L_f
          -4\lambda y^2L_f\right)\,,
    \\
    c_{\phi^5} 
    & = \frac{\kappa \lambda^2 \zeta(3)}{1024\pi^4 \sqrt{T}} \,,
    \\
    r_{\partial^2\phi^3} 
    & = -\frac{\kappa \lambda \zeta(3)}{1536\pi^4 T^{3/2}} \,,
    \\ 
    c_{\phi^6} 
    &= -\frac{7 \zeta(3)}{192 \pi^4} y^6 + \frac{\zeta(3)}{6144 \pi^4} \lambda^3\,,
    \\
    r_{\partial^4\phi^2}
    &= -\frac{7 \zeta(3)}{384 \pi^4 T^2} y^2\,,
    \\
    r_{\partial^2\phi^4} 
    &= \frac{35 \zeta(3)}{576 \pi^4 T} y^4 - \frac{\zeta(3)}{4608 \pi^4 T} \lambda^2\,.
\end{align}
Here we defined
\begin{align}
    L_b
    &\equiv 2\log\frac{\overline{\Lambda}}{4\pi e^{-\gamma_\mathrm{E}}T}\,,
    \\
    L_f
    &\equiv L_b(\overline{\Lambda})+4\log(2)\,,
    \\
    c
    &\equiv \frac{1}{2}\left(
    \log\frac{8\pi}{9}
    +\frac{\zeta'(2)}{\zeta(2)}
    -2\gamma_\mathrm{E}
    \right)\,,
\end{align}
with $\overline{\Lambda}$ being the $\overline{\mathrm{MS}}$ matching scale, $\gamma_\mathrm{E}$ being the Euler-Mascheroni constant and $\zeta(z)$ being the Riemann zeta function. Finally, $\overline\Lambda_\mathrm{3d}$ is the 3D EFT running scale.

We take these matching equations and plug in the running of 4D parameters to remove the matching scale dependence, and fix the reference scale for the running at the critical temperature, $\mu_0 = \pi T_c^{(0)}$ (see \cite{Navarrete:2025yxy} for details). The critical temperature is defined from the condition that $\Delta V = V(\varphi_\mathrm{TV}) - V(\varphi_\mathrm{FV}) = 0$, where $\varphi_\mathrm{TV}$ is the true vacuum state, so $T_c^{(0)}$ is obtained from this equation at $\mathcal{O}(\epsilon^0)$. We additionally carry out a shift in $\phi$ to remove the tadpole term perturbatively through $\mathcal{O}(y^6)$, which fixes $\varphi_\mathrm{FV}=0$. This leaves us with a 3D EFT Lagrangian as a function of modified WCs.

\subsubsection{Numerical evaluation}

We follow the exact same approach as in the previous example to evaluate the statistical rate in this model. In this case, we compute the full nucleation rate $\Gamma = \Gamma_{\rm dyn} \Gamma_{\rm stat}$, in order to estimate the nucleation temperature. We approximate $\Gamma_{\rm dyn}$ by the prefactor given by \texttt{BubbleDet} for the $\mathcal{O}(\epsilon^0)$ action $S_0$ in the no-damping limit.

\begin{figure}[h!]
    \centering
    \includegraphics[width=0.8\linewidth]{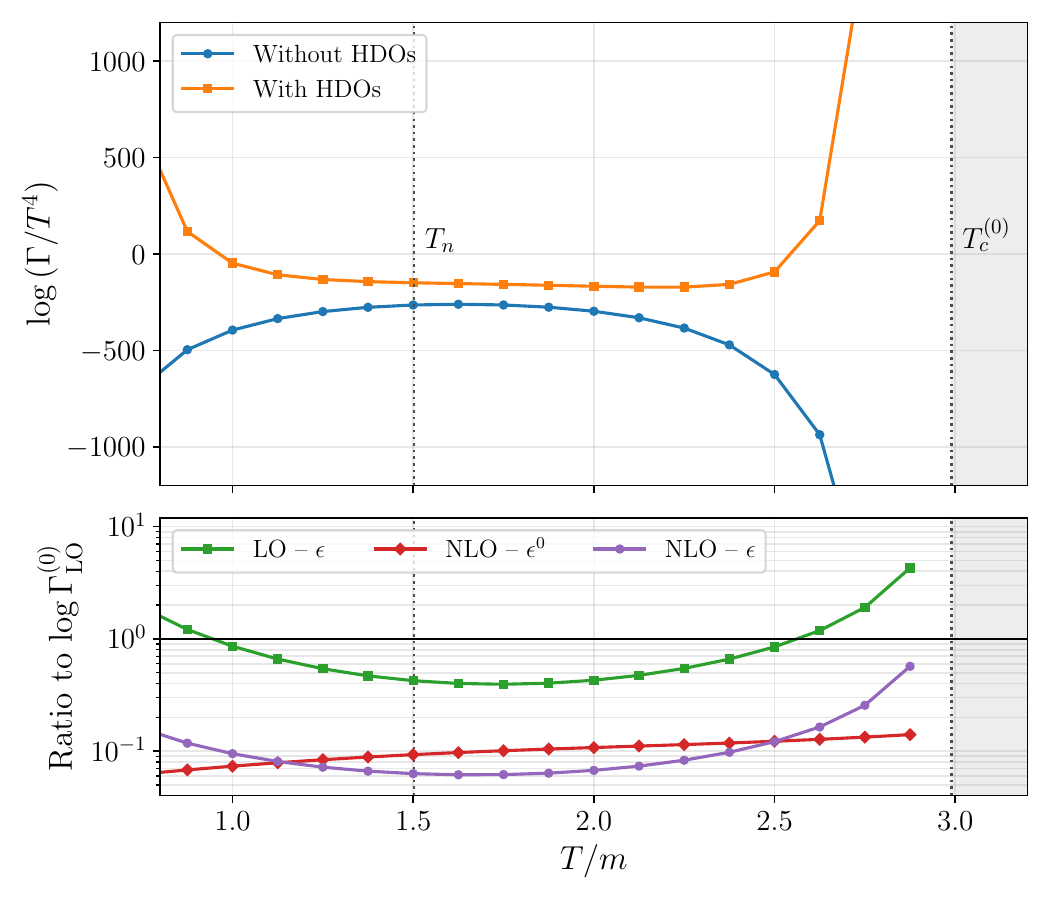}
    \caption{NLO nucleation rate in benchmark point \eqref{eq:ex2 params} with $y=0.9$ for different values of the temperature. Top: full NLO nucleation rate with and without $\mathcal{O}(\epsilon)$ corrections. Bottom: relative contribution of each of the different orders, normalized by $\log\Gamma_{\rm LO}^{(0)} = -S_0[\varphi_0]$. We also show the nucleation temperature, computed from eq.~\eqref{eq:nuctemp}.}
    \label{fig:yukawa plot 1}
\end{figure}

In fig.~\ref{fig:yukawa plot 1} we show $\log \Gamma$ at NLO with and without $\mathcal{O}(\epsilon)$ corrections for different temperatures, at $y=0.9$. We observe that $\log \Gamma^{(0)}$ is similar to the full $\log \Gamma$ for $T$ between $0.4T_c^{(0)}$ and $0.8T_c^{(0)}$. As $T$ decreases, HDO effects naturally become increasingly large, signaling the breakdown of the high-$T$ EFT expansion. On the other hand, close to $T_c^{(0)}$, the $\mathcal{O}(\epsilon^0)$ rate goes to zero, as the vacua become degenerate. Interestingly, the $\mathcal{O}(\epsilon)$ correction also diverges, but positively. Indeed, in this limit, the bounce becomes thin-walled, so it can be approximated by a step function with a growing size $R \sim (\Delta V_0)^{-1}$ and $\varphi_0(r \leq R)\thickapprox\varphi_\mathrm{TV}$, $\varphi_0(r > R) \approx \varphi_{\rm FV}$ \cite{Coleman:1977py}. Right at $T_c^{(0)}$, the radius grows infinite and the bounce tends to $\varphi_\mathrm{TV}$ for all $r$. Therefore, the LO--$\epsilon$ contribution to $\log \Gamma$ asymptotes to
\begin{align}
    S_1[\varphi_0] &\sim \int_x \left[ c_{\phi^5} \varphi_0^5 +  c_{\phi^6} \varphi_0^6 \right] \sim \Delta V_1 R^3\,,
\end{align}
and the contribution of derivative operators is subleading, as they only yield corrections to the surface tension that scale quadratically with the bubble radius. Hence, $\log\Gamma$ diverges, while its sign is controlled by the value of $\Delta V_1$. The two limits signal distinct limitations: the low-$T$ behavior indicates a breakdown of the high-$T$ DR expansion, whereas near $T_c^{(0)}$ the correction to the potential difference between vacua makes the expansion around the leading bounce unreliable, without necessarily invalidating the high-$T$ EFT itself.

In the lower plot in fig.~\ref{fig:yukawa plot 1} we show the different terms contributing to the full NLO $\mathcal{O}(\epsilon)$ rate: at large Yukawa coupling ($y\sim 0.9$) HDOs are more impactful than loop effects. While the loop expansion is well under control, higher EFT orders contribute at the very least a $30\%$ of the total logarithmic nucleation rate. This indicates that even higher orders in the EFT may be relevant at this point in parameter space. The same has been shown to be true in the estimation of critical parameters in \cite{Chala:2025oul,Bernardo:2026whs}.

\begin{figure}[h!]
    \centering
    \includegraphics[width=0.8\linewidth]{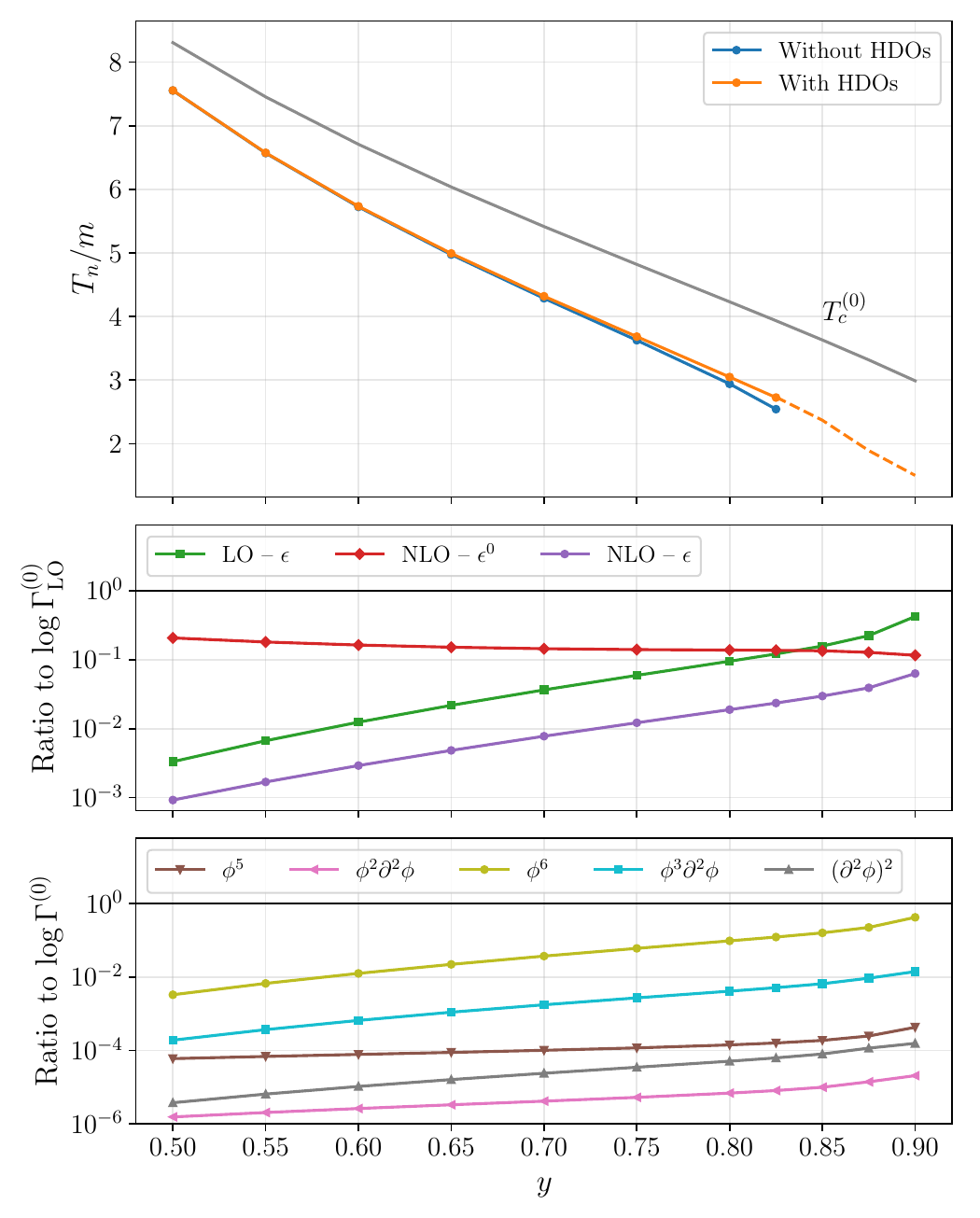}
    \caption{Nucleation temperature ($T_n$) in benchmark point \eqref{eq:ex2 params} for different values of the Yukawa coupling $y$. Top: comparison of the total and $\mathcal{O}(\epsilon^0)$ estimations of $T_n$, along with the $\mathcal{O}(\epsilon^0)$ critical temperature; the dashed line extends the region of parameter space where nucleation occurs when including HDOs. Center: relative contribution of each order to the nucleation rate at $T_n$, normalized by $\log\Gamma_{\rm LO}^{(0)} = -S_0[\varphi_0]$. Bottom: relative contribution of each HDO to the NLO--$\epsilon^0$ nucleation rate at $T_n$.}
    \label{fig:yukawa plot 2}
\end{figure}

In fig.~\ref{fig:yukawa plot 2} we compute the nucleation temperature $T_n$ at different values of $y$. We define $T_n$ from the approximate condition that
\begin{equation}
    \Gamma(T_n) \simeq H(T_n)^4\,,
    \label{eq:nuctemp}
\end{equation}
where $H(T)$ is the Hubble rate. We assume a radiation-dominated universe with one scalar and one Dirac fermion, so that $H(T) = \frac{\pi}{\sqrt{20}} \frac{T^2}{\overline{M}_\mathrm{Pl}}$, where $\overline{M}_\mathrm{Pl} = (8 \pi G)^{-1/2} \simeq 2.435 \times 10^{18}~\mathrm{GeV}$ is the reduced Planck mass.

In~\cite{Navarrete:2025yxy} the authors compare the 3D EFT prediction for the critical latent heat $L(T_c)$ (at superrenormalizable level), with the prediction obtained with a method that does not rely on high-$T$ expansions. The large difference between these two for $y \gtrsim 0.7$, precisely as the transition becomes stronger, signals the breakdown of the high-$T$ approximation. This is also the regime where HDOs in the 3D EFT become increasingly sizable. The reason is that $L(T) \propto d\Delta V / dT$ probes large field values (concretely, $\varphi_\mathrm{TV}$), for which the high-$T$ assumption that $\varphi \ll T$ is no longer satisfied. Analyzing the upper and middle plots in fig.~\ref{fig:yukawa plot 2}, we see that HDOs only moderately modify the prediction for $T_n$ in the $y \gtrsim 0.7$ region, which indicates that the high-$T$ approximation is trustworthy at this order. This can be understood from the fact that in this moderately-strong transition regime the escape point for the transition (largest field value probed by the bounce) is far from the true vacuum. Importantly, however, they do change the region of parameter space in which bubbles can nucleate, extending it beyond $y\approx 0.83$. 

In the bottom plot in fig.~\ref{fig:yukawa plot 2} we isolate the contribution to $\log\Gamma$ from each of the HDOs, evaluated at $T_n$, and observe that $\phi^6$ is much more relevant than derivative terms. A similar conclusion was reached in~\cite{Bernardo:2026tki} for the classically conformal Abelian Higgs model. This result reveals that the naive EFT power counting that treats all HDOs on equal footing in strong PTs is not accurate: derivative insertions are more suppressed than field insertions, and the latter might have to be resummed. This supports the arguments recently presented in \cite{Gould:2026rqz}. Notice that the derivative operator $\phi^3\partial^2\phi$ is actually more relevant than $\phi^5$, but this is due to the fact that dimension-5 operators do not receive genuine fermionic contributions in the matching (their mild $y$ dependence comes from the removal of the tadpole operator).

\section{Conclusions}

In this work, we have developed a systematic perturbative framework for incorporating higher-dimensional operators in the bubble nucleation rate in scalar high-temperature EFTs, concretely in the statistical factor. At first order in the higher-dimensional interactions and NLO in the semiclassical expansion, our formulation consistently includes corrections to the classical action, the bounce profile, the zero-mode Jacobian, the negative eigenvalue and the fluctuation determinant. It therefore extends the estimation of higher EFT corrections beyond equilibrium thermodynamics \cite{Chala:2024xll,Chala:2025oul,Bernardo:2025vkz,Bernardo:2026whs}. The extension to even higher orders than those explored here is conceptually clear, although certainly more computationally expensive.

A central advantage of our approach is its explicit compatibility with EFT power counting. Expanding in the Wilson coefficients isolates the correction at the retained EFT order and avoids introducing an uncontrolled subset of higher-order terms through partial resummation. With this, we have established the invariance of the statistical rate under perturbative field redefinitions, demonstrating that the result is independent of the operator basis when all contributions are included consistently. Our approach is applicable beyond the conventional high-temperature expansion, as long as the nucleation-scale theory is local in derivatives. This might prove useful for studies of nucleation in the strong transition regime, using alternative expansions like the one recently proposed in \cite{Gould:2026rqz}. Finally, we have identified sufficient conditions for the existence and uniqueness of the perturbative bounce corrections, providing a mathematical foundation for the expansion.

We have applied this framework to assess the validity of the high-temperature approximation in the intermediate-strength transition regime of the scalar–Yukawa model. We have determined that while the latent heat cannot be faithfully described in this regime \cite{Chala:2024xll,Chala:2025oul,Navarrete:2025yxy}, the 3D EFT approximation does not show a clear breakdown in the estimation of the nucleation temperature. Still, for the strongest phase transitions in this model, higher-order corrections may open or close regions of parameter space where nucleation can occur, so they should be evaluated. It would be interesting to check the validity of the high-temperature approximation in other theories that allow for stronger transitions, such as the singlet-extended SM \cite{Profumo:2007wc,Espinosa:2011ax,Cline:2012hg,Biekotter:2025npc}. Results similar to ours have been recently obtained in the classically conformal Abelian Higgs model, using a different approach \cite{Bernardo:2026whs,Bernardo:2026tki}. Extending our method to gauge theories, although conceptually similar, involves technical difficulties associated with the appearance of gauge-dependent matrix propagators. Their treatment constitutes the next step towards more precise predictions of cosmological phase transition parameters, and we leave it to a future publication.

\section*{Acknowledgments}

We thank Mikael Chala, Juan Carlos Criado and Adrián Moreno-Sánchez for useful discussions. LG and JLM are supported by the FPU program under grant numbers FPU23/02026 and FPU23/02028, respectively. This work has received funding from the European Research Council under grant agreement n. 101230200, MICIU/AEI/10.13039/501100011033 and ERDF/EU (grants  PID2022-139466NB-C21, PID2022-139466NB-C22 and PID2025-170216NB-I00).
Assistance on several details was provided by Claude Opus 5.5 and GPT Astra 6. 

\appendix
\renewcommand{\thesection}{\Alph{section}}
\renewcommand{\thesubsection}{\Alph{section}.\arabic{subsection}}
\renewcommand{\theequation}{\Alph{section}.\arabic{equation}}

\section{Field redefinition invariance}
\label{app:fr invariance}

By definition, observables in QFT are invariant under field redefinitions (FRs), that is, under invertible changes in the dynamical field variables that we use to describe the system. When dealing with EFTs, and FRs which are not exactly invertible, but perturbative in the EFT expansion parameter, FR invariance can also be proven order by order, provided the perturbative expansion is treated consistently.

Following \cite{Criado:2018sdb}, this section establishes the exact and perturbative FR invariance of both the classical action evaluated at its stationary points, and the full NLO statistical rate.

\subsection{Invariance of the classical action at its stationary points}

Take $\varphi$ to be a stationary point of an action $S$, that is, it satisfies the EOM
\begin{equation}
    S'[\varphi] = 0\,.
\end{equation}
Consider now a change of field variable, \textit{i.e.} $\phi \to \tilde{\phi} = F(\phi)$ where $F(\phi)$ is an invertible function (henceforth we will refer to such transformations as FRs). The redefined action is then simply the composition $\tilde{S} = S \circ F^{-1}$. From here, it trivially follows that $\tilde{S}[\tilde{\phi}] = S[\phi]$. Applying the chain rule for functional derivatives we have that
\begin{equation}
    \frac{\delta \tilde{S}}{\delta \psi(x)} \biggr|_{\psi = \tilde{\varphi}} = \int_y \left(\frac{\delta S}{\delta \psi(y)} \circ F^{-1} \right)\biggr|_{\psi=\tilde{\varphi}} \frac{\delta F^{-1}(\psi(y))}{\delta \psi(x)} \biggr|_{\psi=\tilde{\varphi}} = 0\,,
\end{equation}
for $\varphi = F^{-1}(\tilde{\varphi})$, where $\varphi$ is a stationary point of $S$, \textit{i.e.} $\frac{\delta S}{\delta \psi} \bigr|_{\psi = \varphi} = 0$. This shows that for each stationary point $\varphi$ of an action, there exists a stationary point $\tilde{\varphi}$ of the redefined action, which are in general different, but their corresponding actions are equal.

We can also see this FR invariance order by order if we assume a local, perturbative FR $\phi \to \tilde\phi = \phi + \epsilon f(\phi)$, with $\epsilon \ll 1$ and $f$ a polynomial in $\phi$ and its derivatives. Indeed,
\begin{equation}
    \tilde S[\phi] 
    = 
    S[\phi - \epsilon f(\phi) + \mathcal{O}(\epsilon^2)] 
    = 
    S[\phi] - \epsilon\, S'[\phi] f(\phi) + \mathcal{O}(\epsilon^2) \,,
    \label{eq:FR of S}
\end{equation}
where $S'[\phi] f(\phi)$ is short for $\int_y \frac{\delta S}{\delta \psi(y)}\bigr|_{\psi=\phi}\, f(\phi(y))$. Now, from eq.~\eqref{eq:FR of S}
\begin{equation}
    \tilde{S}'[\phi] = S'[\phi] - \epsilon \left( S''[\phi] f(\phi) + S'[\phi] f'(\phi) \right) \,,
\end{equation}
where $f'(\phi) \equiv \frac{\delta f(\psi(y))}{\delta \psi(x)} \bigr|_{\psi=\phi}$. The function $\tilde\varphi$ satisfying the EOM for $\tilde S$ is $\tilde \varphi = \varphi + \epsilon f(\varphi)$:
\begin{equation}
    \tilde{S}'[\varphi + \epsilon f(\varphi)] = S'[\varphi + \epsilon f(\varphi)] - \epsilon \biggl( S''[\varphi] f(\varphi) + \underbrace{S'[\varphi]}_0 f'(\varphi) + \mathcal{O}(\epsilon) \biggr) = 0 + \mathcal{O}(\epsilon^2)\,.
\end{equation}

A straightforward evaluation at $\tilde\varphi$ in eq.~\eqref{eq:FR of S} yields the same action
\begin{equation}
     \tilde S[\tilde \varphi] = S[\varphi + \epsilon f(\varphi)] - \epsilon S'[\varphi] f(\varphi) + \mathcal{O}(\epsilon^2) = S[\varphi] + \mathcal{O}(\epsilon^2) \,.
\end{equation}

\subsection{Invariance of the NLO statistical rate}

We will now show that, while the functional determinant of the fluctuation operator $\delta^2 S/\delta \phi^2$ does change under a FR, this difference is compensated by its Jacobian at the level of the path integral measure, ensuring that the nucleation rate remains FR invariant. 

The NLO statistical rate is given by the exponential of the quantum effective action at a stationary point of the action (see eq.~\eqref{eq:Seff}). Assuming a FR of the form $\phi \to \tilde{\phi} = F(\phi)$, the second functional derivative of the redefined action is
\begin{multline}
    \frac{\delta^2 \tilde{S}}{\delta \psi(y) \delta \psi(x)}\biggr|_{\psi=\tilde{\varphi}}
    = \int_t \int_z \left( \frac{\delta^2 S}{\delta \psi(t) \delta \psi(z)} \circ F^{-1} \right) \frac{\delta F^{-1}(\psi(t))}{\delta \psi(y)} \frac{\delta F^{-1}(\psi(z))}{\delta \psi(x)} \biggr|_{\psi=\tilde{\varphi}} 
    \\[1ex]
    + \int_z \left(\frac{\delta S}{\delta \psi(z)} \circ F^{-1} \right) \frac{\delta^2 F^{-1}(\psi(z))}{\delta \psi(y) \delta \psi(x)}\biggr|_{\psi=\tilde{\varphi}}\,.
\end{multline}
Therefore, at a stationary point the second term vanishes and we are left with an object that transforms as a tensor under FRs, $\tilde{S}''[\tilde{\varphi}] = J^\mathrm{T} S''[\varphi] J$, where we have introduced matrix notation and the Jacobian matrix $J = \delta F^{-1}/\delta \psi\bigr|_{\psi=\tilde\varphi}$.

The nucleation rate is related to the quantum effective action evaluated at stationary points $\varphi$ of the classical action $S$. At the one-loop level:
\begin{equation}
    e^{-S_\mathrm{eff}[\varphi]} = \int \mathcal{D}\eta\, e^{-S[\varphi+\eta]} \thickapprox e^{-S[\varphi]} \int \mathcal{D}\eta\, e^{-\frac{1}{2} \eta S''[\varphi] \eta}\,.
\end{equation}
In the field-redefined theory, we instead compute
\begin{align}
    e^{-\tilde{S}_\mathrm{eff}[\tilde\varphi]} &= \int \mathcal{D}\xi\, e^{-\tilde S[\tilde\varphi+\xi]} \thickapprox e^{-\tilde S[\tilde \varphi]} \int \mathcal{D}\xi\, e^{-\frac{1}{2} \xi \tilde S''[\tilde\varphi] \xi} = e^{-S[\varphi]} \int \mathcal{D}\xi\, e^{-\frac{1}{2} \xi J^\mathrm{T} S''[\varphi] J \xi} \nonumber\\
    &= e^{-S[\varphi]} \det (J)^{-1} \int \mathcal{D}\eta\, e^{-\frac{1}{2} \eta S''[\varphi] \eta} \,,
\end{align}
where we have applied the change of variables $\eta = J \xi$ in the last line. Therefore, there is a mismatch between the two expressions due to the Jacobian associated to the FR. This is not surprising: a change of field variables at the level of the path integral only leaves the quantum effective action invariant when consistently including the corresponding change of the integration measure \cite{Criado:2018sdb,Falls:2018olk}. This can also be seen at the level of the fluctuation determinant, since $\det(\tilde S''[\tilde \varphi]) = \det(J)^2 \det(S''[\varphi])$.

In DimReg, however, we need not keep track of the Jacobian when dealing with perturbative FRs, as in this case $\det(J) = 1$. First consider the same FR as before, $\tilde\phi = \phi + \epsilon f(\phi)$, so
\begin{equation}
    \det(J) = \det(1 - \epsilon g_\varphi + \mathcal{O}(\epsilon^2)) \thickapprox 1 - \epsilon \,\tr(g_\varphi) \,, 
    \label{eq:coson}
\end{equation}
where we have defined $g_\varphi \equiv f'(\varphi)$. Let us prove that in DimReg the trace of $g_\varphi$ is identically zero if $f(\varphi)$ is a polynomial in $\varphi$ and its derivatives. The kernel of $g_\varphi$ must be of the form
\begin{equation}
    g_\varphi(x,y) = P(\varphi(x), \partial_i \varphi(x) \partial_i, \partial^2, \dots) \delta^D(x-y) \,,
\end{equation}
where $P$ is a polynomials and the unpaired derivatives act on the right. 
The trace can then be written as follows:
\begin{align}
    \tr(g_\varphi) &=  \int_x \int_y \delta^D(x-y) g_\varphi(x,y) = \int_x \lim_{y \to x} P(\varphi(x), \partial_i \varphi(x) \partial_i, \partial^2, \dots)  \delta^D(x-y) \nonumber\\
    &= \int_x \lim_{y \to x} P(\varphi(x), \partial_i \varphi(x) \partial_i, \partial^2, \dots)  \int_p e^{i p \cdot (x-y)} \,,
\end{align}
where we have introduced the momentum space representation of the delta with $\int_p \equiv \int \frac{d^D p}{(2\pi)^D}$. All unpaired derivatives acting on the delta will yield a polynomial in $p$ of the form
\begin{align}
    &\int_x \lim_{y \to x} \int_p P(\varphi(x), \partial_i \varphi(x) p_i, p^2 \dots) e^{i p \cdot (x-y)} \nonumber\\
    &= \int_x \int_p P(p; \varphi(x), \partial_i \varphi(x) \partial_i \varphi(x), \partial^2 \varphi(x) \dots)\,,
\end{align}
where $P$ is now a polynomial in momentum and fields and their derivatives. In DimReg, all power-divergent integrals of the form
\begin{equation}
    \int_p p^\alpha = 0 \,.
    \label{eq:vanishing dimreg}
\end{equation}
Therefore, the trace above vanishes identically. This is the path integral formulation of the argument developed in \cite{tHooft:1973wag, Criado:2018sdb}, which shows this cancellation diagrammatically in DimReg by expressing the Jacobian as ghost action with unit propagator.

\section{On the existence of perturbative bounce solutions}
\label{app:existence}

We have assumed throughout that the eq.~\eqref{eq:diffeq phi1} with boundary conditions \eqref{eq:bcs phi1} has an $\mathcal{O}(D)$-symmetric solution. Should this not be the case, we could not find a (perturbative) bounce solution to $S = S_0 + \epsilon S_1$. We can gain some insight into the solvability of this partial differential equation by means of the Fredholm alternative theorem \cite{Fredholm1903}. In particular, calling
\begin{align}
    Q &\equiv -\frac{d^2}{dr^2} - \frac{D-1}{r} \frac{d}{dr} + V''(\varphi_0(r)) \,,
    \\
    q &\equiv -S_1'[\varphi_0(r)]\,,
\end{align}
we will address the existence of solutions to 
\begin{equation}
    Q\phi(r) = q(r)\,,\qquad \text{with } \dot\phi(0) = \lim_{r\to\infty}\phi(r) = 0\,.
    \label{eq:second order ode}
\end{equation}
Let $z(r)$ be a zero mode of $Q$, that is, a solution to the homogeneous equation
\begin{equation}
    Qz(r) = 0\,,\qquad \text{with } \dot z(0) = \lim_{r\to\infty}z(r) = 0\,.
    \label{eq:zero modes equation}
\end{equation}
Fredholm alternative theorem states that exactly one of the following holds:

\begin{enumerate}
    \item There is no such $z(r)\neq 0$. Then,~\eqref{eq:second order ode} has a \textit{unique solution}.

    \item There exists $z(r)\neq 0$. Then,~\eqref{eq:second order ode} has \textit{no solution}, unless $q$ and $z$ are orthogonal, \textit{i.e.}, $\int_0^\infty dr\, r^{D-1} q(r) z(r) = 0$ for all $z(r)$, in which case the solution exists and is nonunique.
\end{enumerate}

Notice that not only does this give a sufficient condition for $\varphi_1(r)$ to exist, but also for every correction $\varphi_i(r)$, since their differential equations read, in general
\begin{equation}
    Q\varphi_i(r) = q_i(r)\,,\qquad \text{with } \dot{\varphi}_i(0) = \lim_{r\to\infty}\varphi_i(r) = 0\,.
\end{equation}
Therefore, the proof of existence of $\varphi_1$ trivially generalizes to the existence of a complete power series solution to $S'[\phi] = 0$.

In light of the above, we study in which cases $Qz_a(r) = 0$ admits non-zero $z_a(r)$. Since $Q$ is the radial ($\ell=0$) sector of the differential operator $S_0''[\varphi_0]$, we equivalently study the existence of radial zero modes. Crucially, the analysis of zero modes of the elliptic operator $S_0''[\varphi_0]$ and the existence of the LO bounce solution are  connected~\cite{ADACHI20181121, Lewin2020}. Broadly speaking, the main result can be summarized as follows: \textit{if a bounce solution exists and is unique, then $Q$ has no zero modes, and, therefore, the perturbative corrections exist and are also unique}.

More precisely, let $0 < \varphi_{\rm B} < \varphi_{\rm TV}$, and $V(\phi)$ be a sufficiently differentiable real function satisfying the following conditions:
\begin{flalign*}
    \qquad \mathrm{(C1)} \qquad & V'(0) = 0\,\quad\mathrm{and}\quad V''(0) > 0 \,, &\\
   \qquad  \mathrm{(C2)} \qquad &V'(\phi) > 0\,  \text{ for }\, 0 < \phi < \varphi_\mathrm{B}\,, &\\
  \qquad  \mathrm{(C3)} \qquad & V'(\varphi_{\rm B}) = 0\,\quad\mathrm{and}\quad V''(\varphi_{\rm B}) < 0 \,, &\\
   \qquad  \mathrm{(C4)} \qquad & V'(\phi)<0\,  \text{ for }\,  \varphi_{\rm B} < \phi < \varphi_{\rm TV}\,, &\\
   \qquad \mathrm{(C5)} \qquad & V'(\varphi_{\rm TV}) = 0\,\quad\mathrm{or}\quad \varphi_{\rm TV} = \infty \,. &
\end{flalign*}
These imply the existence of a barrier at $\varphi_{\rm B}$. In particular, it comprises two cases: (i) the existence of two minima at $\varphi_{\rm FV} = 0$ and $\varphi_{\rm TV}$; and (ii) the existence of just one minimum at $\varphi_{\rm FV} = 0$ and a runaway potential after the barrier. Let us state two more conditions:
\begin{flalign*}
    \qquad  \mathrm{(C6)} \qquad & \phi V''(\phi) \leq V'(\phi) \text{ for } 0<\phi<\varphi_{\rm B}\,. &
    \\
    \qquad  \mathrm{(C7)} \qquad & \phi V''(\phi) = \eta V'(\phi) \text{ has exactly one solution in } (\varphi_{\rm B}, \varphi_{\rm TV}),~~\forall \,\eta > 1\,. &
\end{flalign*}
If (C1)--(C7) are met, then the existence of a bounce solution guarantees that $S_0''[\varphi_0]$ has exactly $D$ zero modes, corresponding to the translational ones $\partial_i\varphi_0(x)$, lying at the $\ell=1$ sector~\cite{Lewin2020, ADACHI20181121}. Namely, $Q$ has no zero modes. Therefore, by the case 1 of Fredholm alternative theorem, if the LO bounce solution exists, so does its perturbation $\varphi_1$. Conditions (C1)--(C7) also ensure that the bounce solution, if it exists, is unique.

Ref.~\cite{ADACHI20181121} gives some deeper insight about the existence of the bounce solution itself. Assuming (C1), together with (C8) and (C9):
\begin{flalign*}
    \qquad  \mathrm{(C8)} \qquad & V(\phi) < V(0) \text{ for some } \phi > 0 \,, & \\
    \qquad  \mathrm{(C9)} \qquad & \lim_{\phi\to\infty} \frac{V'(\phi)}{\phi^{\frac{D+2}{D-2}}} \geq 0 ~~(\text{for } D \geq3) \,, &
\end{flalign*}
they prove that at least one bounce solution exists. Notice that this is an alternative and independent proof of existence to the one found by Coleman, Glaser and Martin (CGM), in~\cite{Coleman:1977th}. In particular, (C9) is changed by the condition that there exist suitable $a,b,\alpha,\beta > 0$ such that $\alpha < \beta < 2D/(D-2)$ and 
\begin{flalign*}
   \qquad  \mathrm{(CGM)} \qquad & V(\phi) \geq a |\phi|^\alpha - b |\phi|^\beta\,. &
\end{flalign*}
These two formulations are an alternative set of hypotheses in the sense that neither implies the other. As opposed to \cite{Coleman:1977th}, in~\cite{ADACHI20181121} they give an existence and uniqueness theorem as well. Assuming the complete set (C1)--(C9), the bounce exists and is unique. In addition, if $\varphi_{\rm TV} < \infty$, and (C8), (C9) are replaced by the stronger
\begin{flalign*}
    \qquad  \mathrm{(C8')} \qquad & V(\varphi_{\rm TV}) < V(0) \,, & \\
    \qquad  \mathrm{(C9')} \qquad & V'(\phi) > 0 \text{ for } \phi > \varphi_{\rm TV} \,, &
\end{flalign*}
then the bounce solution exists, it is unique, and it is bounded from above by $\varphi_{\rm TV}$. 

To summarize, given a potential $V(\phi)$ satisfying (C1)--(C7), together with any sufficient condition for the existence of a bounce solution, \textit{e.g.}, (C8)--(C9) or (CGM), there exists a (perturbative) power series solution in $\epsilon$ to $S'[\phi]=0$, where $S[\phi]=\sum_{k\geq0} \epsilon^k S_k[\phi]$ and $S_0''[\phi] = -\partial^2 + V''(\phi) $. 

Let us note that the discussion above establishes a formal perturbative solution, but does not by itself establish the convergence of the series to an exact bounce solution. Under suitable analyticity and invertibility assumptions, an analytic implicit function theorem could establish a local solution branch represented by this series. Verifying these assumptions, particularly in the presence of HDOs, and determining whether the resulting radius of convergence includes the physical point $\epsilon=1$ (as we insert $\epsilon$ as a formal parameter to organize the expansion) are beyond the scope of this work. Convergence is not, however, necessary for a finite-order expansion to be useful: an asymptotic expansion may also yield accurate approximations. Comparing successive $\mathcal{O}(\epsilon^n)$ corrections provides a practical check of the perturbative hierarchy, even if it does not establish convergence or a bound on the truncation error.

\subsubsection*{A direct approach to the solvability condition}

While these solvability arguments rely on the Fredholm alternative theorem, we can alternatively understand this by working out a solution by quadratures. First, we study the solutions to the homogeneous equation $Q\phi(r) = 0$. Since $\varphi_0(r)$ is approximately constant as $r\to0$ and $r\to\infty$, the leading behavior of the differential equation is
\begin{equation}
    -\ddot\phi(r) - \frac{D-1}{r} \dot\phi(r) + M^2 \phi(r) = 0\,,
    \label{eq:ode at asymptotics}
\end{equation}
where $M^2=\omega^2\equiv V''(0)$ at $r\to\infty$, and $M^2=\kappa^2\equiv V''(\varphi_0(0))$ at $r\to 0$. We consider $\omega^2 > 0$, as required by condition (C1). The asymptotic behaviors are:
\begin{equation}
\begin{array}{rll}
   \phi(r)  &  \sim r^{-\frac{D-1}{2}} \left (A_1 e^{\omega r} + A_2  e^{-\omega r} \right) & \quad\mathrm{as}\quad r\to\infty \,, \\[0.5em]
    \phi(r) & \sim B_1 \left[ 1+\frac{\kappa^2}{2D} r^2 \right] + B_2 \, r^{2-D} &\quad\mathrm{as}\quad r\to0 \,. 
\end{array}
\label{eq:asymptotics}
\end{equation}
Let $u(r),v(r)$ be two independent solutions to the homogeneous equation, such that $u(r)$ is regular at the origin ($B_2=0$) and $v(r)$ decays to zero at infinity ($A_1=0$). We will suppose that the potential is such that there exists no function satisfying both endpoint conditions (otherwise, the solution would be a zero mode of $Q$). It can be verified that a particular solution $h(r)$ to the inhomogeneous equation~\eqref{eq:second order ode} takes the following form:
\begin{equation}
    h(r) = u(r) \int_0^r d \rho \, \frac{v(\rho) q(\rho)}{W(\rho)} - v(r) \int_0^r d \rho \, \frac{u(\rho) q(\rho)}{W(\rho)} \,,
\end{equation}
where $W(r)=u(r)\dot v(r)-\dot u(r)v(r)$ is the Wronskian. We can find an explicit form for $W(r)$:
\begin{equation}
    \dot W(r) = u(r) \ddot v(r) - \ddot u(r) v(r) = -\frac{D-1}{r} W(r) \, \implies \, W(r) = \frac{K}{r^{D-1}}\,,
\end{equation}
where we used that $u(r),v(r)$ satisfy the homogeneous equation and defined a nonzero constant $K$ fixed by the potential contained in $Q$. Altogether,
\begin{equation}
    h(r) = \frac{1}{K} \biggl[ u(r) \int_0^r d \rho \, \rho^{D-1} v(\rho) q(\rho) - v(r) \int_0^r d \rho \, \rho^{D-1} u(\rho) q(\rho) \biggr]\,.
    \label{eq:particular solution}
\end{equation}

Thus, the solution $\varphi_1(r)$ satisfying the specific BCs in~\eqref{eq:second order ode} must be a linear combination
\begin{equation}
    \varphi_1(r) = C_1 u(r) + C_2 v(r) + h(r)\,.
\end{equation}
Notice that we require convergence of $\varphi_1(r)$ at infinity, but $u(r)$ diverges there. Therefore, $C_1$ must be fine-tuned to cancel the behavior of $h(r)$ at infinity. Similarly, in order to balance the divergence of $\dot v(r)$ at $r\to 0$, as required by $\dot \varphi_1(0)=0$, $C_2$ must be fixed accordingly. It follows
\begin{equation}
    C_1 = -\frac{1}{K}\int_0^\infty d\rho\, \rho^{D-1} v(\rho) q(\rho)\,,
    \qquad
    C_2 = 0\,.
\end{equation}
In order to obtain these values, note that $q(r)$ must decay rapidly enough at infinity. Therefore, $\varphi_1(r)$ can be written as follows
%\footnote{An identical result is obtained via Green's function construction.}:
%
\begin{equation}
      \varphi_1(r) = -\frac{1}{K} \biggl[ u(r) \int_r^\infty d \rho \, \rho^{D-1} v(\rho) q(\rho) + v(r) \int_0^r d \rho \, \rho^{D-1} u(\rho) q(\rho) \biggr]\,.
\end{equation}

We obtained a solution to the differential equation which is unique. For that, we had to assume that no solution to the homogeneous equation is both regular at the origin and decaying at infinity, or, equivalently, that no zero mode of $Q$ exists. Should such a function exist, call this function $v(r)=z(r)$. Then, $u(r)$ diverges at both endpoints. The previous argument would have broken down when trying to determine $C_2$. Indeed, after obtaining $C_1$, at $r\to 0$,
\begin{equation}
    \dot \varphi_1(r)  \sim -\frac{1}{K} \dot u(r) \int_0^\infty d\rho \rho^{D-1} z(\rho) q(\rho)
\end{equation}
is divergent, unless the integral $I\equiv \int_0^\infty d\rho \rho^{D-1} z(\rho) q(\rho) = 0$. Therefore, no choice of $C_2$ can make $\varphi_1(r)$ satisfy the boundary conditions, unless $I=0$, in which case $z(r)$ and $q(r)$ are orthogonal. Thus, we recover all the cases in Fredholm alternative theorem.

\section{Solving the propagator equation}
\label{app:propagator}

Our goal is to solve numerically eq.~\eqref{eq:spherical harmonic propagator tilde equation} for each $\ell$. Following \cite{Ekstedt:2022tqk}, for the special case $\ell=1$ we solve the propagator equation shifted by an infrared regulator $s$,
\begin{equation}
    \left[-\partial_r^2 - \frac{D-1}{r} \partial_r + \frac{D - 1}{r^2} + V_0''(\varphi_0(r)) + s \right] G^{s}_{\ell=1}(r, r') = \frac{1}{r^{D-1}}\delta(r-r')\,,
    \label{eq:shifted equation}
\end{equation}
and recover $\widetilde{G}_{\ell=1}(r,r')$ by taking the limit
\begin{equation}
    \widetilde{G}_{\ell=1}(r,r') = \lim_{s \to 0} \left[ G^{s}_{\ell=1}(r, r') - \frac{1}{s} \frac{\Omega_{D-1}}{D S_0[\varphi_0]} \dot \varphi_0(r) \dot \varphi_0(r') \right]\,.
    \label{eq:shifted limit}
\end{equation}
This allows us to solve the differential equation in the $\ell=1$ sector using the same matching conditions at $r=r'$ (due to the delta source term) as in the $\ell \neq 1$ sector. These matching conditions impose the continuity of $\widetilde{G}_\ell(r,r')$ at $r=r'$, and a jump in its first derivative. If $\widetilde{G}^>_\ell(r,r')$ is the solution for $r > r'$ and $\widetilde{G}^<_\ell(r,r')$ is the solution for $r < r'$, we impose
\begin{align}
    &\left[ \widetilde{G}^>_\ell(r,r') - \widetilde{G}^<_\ell(r,r') \right]_{r=r'} = 0 \,, \\
    &\left[ \partial_r \widetilde{G}^>_\ell(r,r') - \partial_r \widetilde{G}^<_\ell(r,r') \right]_{r=r'} = -\frac{1}{(r')^{D-1}} \,,
\end{align}
where the second condition follows from integrating the differential equation across $r=r'$.

We must fix additional boundary conditions for the propagator. Since $V''(\varphi_0(r))$ is smooth at $r=0$ (the bounce is), regularity at the origin imposes that each $\widetilde{G}_\ell(r,r')$ be regular at the origin
\begin{equation}
    \widetilde{G}_\ell(r,r') \sim r^\ell\,, \quad r'>0\,,\, r \to 0\,.
    \label{eq:bc origin}
\end{equation}
In practice, we factor out this known behavior as $\widetilde G_\ell(r,r') = r^\ell u_\ell(r,r')$, where $u_\ell(r,r')$ is finite at $r=0$ and it satisfies
\begin{equation}
    \partial_r u_\ell(r,r') \Bigr|_{r=0} = 0\,.
\end{equation}
On the other hand, if the background approaches the FV as $r \to \infty$, so $V''(\varphi_0(r)) \to \omega^2$. In this limit eq.~\eqref{eq:spherical harmonic propagator tilde equation} can be solved analytically. The decaying homogeneous solution is
\begin{equation}
    \widetilde{G}_\ell(r,r') \sim \frac1{r^{(D-2)/2}} K_{\ell + \frac{D-2}{2}}(\omega r) \sim \frac{1}{r^{(D-1)/2}} e^{-\omega r}\,.
\end{equation}
where $K_\nu(z)$ is the modified Bessel function of the second kind. Therefore $\widetilde{G}_\ell(r,r')$ vanishes as $r\to\infty$, and does so exponentially.

\paragraph{WKB approximation.} While eq.~\eqref{eq:spherical harmonic propagator tilde equation} cannot be solved analytically, in the WKB approximation, however, one can obtain an analytic expression for $\widetilde{G}_\ell(r,r')$ in the large-$\ell$ limit. Closely following the derivation in appendix D in \cite{Ekstedt:2022tqk} (see references therein), we perform a change of variables $r=r_0 e^x$. Then eq.~\eqref{eq:spherical harmonic propagator tilde equation} (for $r\neq r'$) becomes
\begin{equation}
    \Psi''(x) - \eta(x) \Psi(x) = 0,
\end{equation}
with $\widetilde{G}_\ell(r,r') \equiv e^{\frac{2-D}{2} x} \Psi(r=r_0 e^x)$. Here, we have further defined $\eta(x) \equiv r_0^2 e^{2x} W(x) + \bar{\ell}^2$, with $W(x) \equiv V''_0(\varphi_0(r_0 e^x))$ and $\bar{\ell} \equiv \frac{D+2\ell-2}{2}$. This is the equation that one solves employing the WKB approximation in the large-$\ell$ limit. Here we only need the propagator in the coincident limit ($r=r'$), so we impose it in eq.~(D.6) in \cite{Ekstedt:2022tqk} (fixing a typo in that work), which yields
\begin{equation}
    \widetilde{G}_\ell^\mathrm{(WKB)}(r) = \frac{1}{D+2\ell-2} r^{2-D} -  \frac{2}{(D+2\ell-2)^3} r^{4-D} V''_0(\varphi_0(r)) \,.
    \label{eq:WKB approximation}
\end{equation}

\section{Convergence of the radial integrals $\widetilde{T}_\ell$}
\label{app:radial integrals}

In this appendix we prove that the radial integrals $\widetilde{T}_\ell \equiv \widetilde{T}_{\mathrm{b}, \ell} - T_{\mathrm{FV}, \ell}$ in Eq.~\eqref{eq:trace general bounce} are convergent for every $\ell$, so that the divergence is only encountered in the infinite sum. We show this by a detailed study of the small and large $r$ asymptotics. Without loss of generality, we set $\varphi_{\rm FV}=0$ for the discussion. 

First, we state the asymptotics of the bounce. Since $\dot \varphi_0(0) = 0$, then $\varphi_0(r) \sim \varphi_0(0) + \mathcal{O}(r^2)$ when $r \to 0$. Moreover, by imposing $\varphi_0(r)\to 0$ as $r\to\infty$, and analyticity of $V(\phi)$ at the origin, $\varphi_0(r)$ satisfies the differential equation in~\eqref{eq:ode at asymptotics} at infinity, with $M=\omega$. Therefore,
\begin{equation}
    \begin{array}{rll}
   \varphi_0(r)  &  \sim \varphi_0(0) + \mathcal{O}(r^2) & \quad\mathrm{as}\quad r\sim0 \,, \\[0.5em]
    \varphi_0(r) & \sim r^{-\frac{D-1}{2}} e^{-\omega\,r} &\quad\mathrm{as}\quad r\sim\infty \,. 
\end{array}
\end{equation}
Notice that, by the discussion in the appendix~\ref{app:existence}, $\varphi_1(r)$ has the same asymptotics at $r\sim0$, and $\varphi_1(r)\sim r^a e^{-b r}$ at $r\to\infty$, where $a,b$ ($b>0$) depend on the explicit form of $q(r)$. 

Regarding $\mathcal{A}(r), \mathcal{B}(r), \mathcal{C}(r)$ in eq.~\eqref{eq:trace general bounce}, their only explicit dependence on $r$ comes from the expansion of the radial Laplacian. However, these can be removed by means of the EOM of $\varphi_0(r)$, so that all the $r$-dependence in $\mathcal{A}(r), \mathcal{B}(r), \mathcal{C}(r)$ can possibly come from $\varphi_0(r), \dot \varphi_0(r)$ and $\varphi_1(r)$, yielding similar asymptotic behaviors to those of $\varphi_0(r)$ and $\varphi_1(r)$. The only exception is $\mathcal{A}(r)$, which will be shifted by a possibly nonzero constant $\mathcal{A}_{\rm FV}$ at $r\to\infty$.

\paragraph{Limit $r \to 0$.} In this limit eq.~\eqref{eq:spherical harmonic propagator tilde equation} becomes the equation for a homogeneous Euclidean scalar propagator with constant mass $\kappa^2 \equiv V''_0(\varphi_0(0))$, whose solution takes the form
\begin{equation}
    \widetilde{G}_\ell(r) = \frac1{r^{D-2}} I_{\ell + \frac{D-2}{2}}(\kappa r)\, K_{\ell + \frac{D-2}{2}}(\kappa r) \,.
    \label{eq:G flat}
\end{equation}
Its asymptotic behavior as $r \to 0$ is (for $D > 2$):
\begin{equation}
    \widetilde{G}_\ell(r) \propto \frac{1}{r^{D-2}}\,,
    \qquad
    \dot{\widetilde{G}}_\ell(r) \propto \frac{1}{r^{D-1}} \,.
\end{equation}
Plugging in these asymptotics we see that the integrand of $\widetilde{T}_{\mathrm{b},\ell}$ is finite at $r\to 0$. Schematically,
\begin{equation}
    r^{D-1} \left[ \mathcal{A}(r) \frac{1}{r^{D-2}} + \mathcal{B}(r) \frac{1}{r^{D-1}} + \delta_{\ell 1} \mathcal{C}(r) \right] \sim \text{finite} \qquad \text{ as } r\to0\,.
\end{equation}
Similarly, the integrand of $T_{\mathrm{FV}, \ell}$ is finite as well. This proves that the radial integral at each fixed $\ell$ is not UV divergent.

\paragraph{Limit $r\to\infty$.} Since $V''_0(\varphi_0(r)) \sim \omega^2 + \mathcal{O}(e^{-\mu_1 r}),\, \mu_1>0$, eq.~\eqref{eq:spherical harmonic propagator tilde equation} asymptotes to the FV equivalent plus exponentially suppressed corrections. Therefore
\begin{equation}
    \widetilde{G}_\ell(r) \sim G_{\mathrm{FV}, \ell}(r) \left( 1 + \mathcal{O}(e^{-\mu_2 r}) \right)\,, \quad \mu_2 > 0\,,
\end{equation}
and therefore its radial derivatives are also different by an exponential factor. 

Recalling the exponential suppression of $\mathcal{B}(r)$ and $\mathcal{C}(r)$ at infinity, and that $\mathcal{A}(r)\sim \mathcal{A}_{\rm FV}$, we have that
\begin{align}
\notag
    \widetilde{T}_\ell = \widetilde{T}_{\mathrm{b}, \ell} - T_{\mathrm{FV}, \ell} &\sim \int_R^\infty dr r^{D-1} \biggl\{\Bigl[ (1+\mathcal{O}(e^{-\mu_2r})) \mathcal{A}(r)-\mathcal{A}_{\rm FV}\Bigr] G_{\mathrm{FV},\ell}(r)  + \mathcal{O}(e^{-\mu_3r}) \biggr\} \nonumber
    \\
    & \sim \int_R^\infty dr r^{D-1} \mathcal{O}(e^{-\mu_4r}) <\infty \,,
\end{align}
for some positive $\mu_i$ and a lower cutoff radius $R$. Thus, the difference with the FV contribution cuts off any potential IR divergence in the radial integral, making this limit finite as well.

{\small
%%%%%%%%%%%%%%%%%%%%%%%%% BIBLIO %%%%%%%%%%%%%%%%%%%%%%%%%%%%%%%%%%%%%%%%%
%
\bibliographystyle{utphys}
\bibliography{references}

}

\end{document}